\newif\ifarxiv
\arxivtrue % arxivfalse/arxivtrue to change from standard JFM to arXiv-ready submission

\ifarxiv
\documentclass[]{jfm-arxiv}
\else
\documentclass[lineno]{jfm}
\fi

\usepackage{graphicx}
\usepackage{newtxtext}
\usepackage{newtxmath}
\usepackage{natbib}
\usepackage{amsmath}
\usepackage{upgreek}
\usepackage{comment}
\usepackage{textcomp}
\usepackage{pifont}
\usepackage{hyperref}
\usepackage{booktabs}
\usepackage[dvipsnames]{xcolor}
\definecolor{myCyan}{rgb}{0.0, 1.0, 1.0}
\pdfoutput=1

\newcommand{\RomanNumeralCaps}[1]
\linenumbers
\newcommand{\comm}[1]{}
\newcommand{\q}[1]{\textquotedblleft #1\textquotedblright}

\shorttitle{Rise regimes of moderate-viscosity-ratio droplets}
\shortauthor{P. Shi et al.}

\title{Rise regimes of freely rising droplets with a moderate viscosity ratio}

\author{
Pengyu Shi\aff{1,2,3}\corresp{\email{pengyu.shi@ulb.be}},
Dirk Lucas\aff{3},
Jie Zhang\aff{4},
\'{E}ric Climent\aff{2}
\and Dominique Legendre\aff{2,5}
	\corresp{\email{dominique.legendre@imft.fr}}}
	
\affiliation{
\aff{1}Department of Transfers, Interfaces, and Processes, Universit\'e Libre de Bruxelles, 1050 Brussels, Belgium
\aff{2}CNRS, Universit\'e de Toulouse; Toulouse INP; IMFT (Institut de M\'ecanique des Fluides de Toulouse), F-31400 Toulouse, France
\aff{3}Helmholtz-Zentrum Dresden--Rossendorf, Institute of Fluid Dynamics, 01328 Dresden, Germany
\aff{4}State Key Laboratory for Strength and Vibration of Mechanical Structures, School of Aerospace, Xi'an Jiaotong University, Xi'an, PR China
\aff{5}Institut universitaire de France, Paris 75005, France
}

\begin{document}
\maketitle

\begin{abstract}
The dynamics of buoyant droplets rising freely in a large body of an immiscible liquid is investigated numerically for a moderate drop-to-fluid viscosity ratio $\mu^\ast$. We focus on toluene droplets rising in clean water, for which $\mu^\ast=0.62$, and vary the radius over $0.5\,\text{mm}\leq R\leq3.0\,\text{mm}$. Direct numerical simulations are performed in imposed axisymmetric and fully three-dimensional configurations. As $R$ increases, the system displays a rich sequence of rise regimes. Starting from steady vertical rise with an axisymmetric disturbance flow, it first undergoes an internal flow instability associated with an azimuthal mode $m=2$, leading to a biplanar-symmetric wake and reduced terminal speed. This state is followed by a steady oblique regime, in which the $m=1$ mode also becomes unstable and coexists with the $m=2$ mode. At larger radii, the path becomes nearly vertical again before the flow enters an $m=2$ rotating-wave regime, where the wake drifts azimuthally at an approximately constant angular velocity. For still larger droplets, persistent shape oscillations and vortex shedding lead to fully three-dimensional chaotic paths. Simulations initialised from finite-amplitude asymmetric states further reveal several multistable size ranges, in which distinct terminal states coexist depending on the initial condition. Taken together, these findings show that the path instability of moderate-viscosity-ratio droplets differs fundamentally from that of bubbles and solid particles: in most regimes encountered here, axisymmetry breaking is initiated within the droplet, highlighting the central role of the internal flow instability in shaping the subsequent wake structure, rise speed and droplet dynamics.

\end{abstract}


\section{Introduction}
In our previous study (\citealp*{2025_Shi_drop}), we investigated numerically the stability of a uniform flow past a fixed spherical droplet over a wide range of Reynolds numbers and drop-to-fluid viscosity ratios $\mu^\ast$. A remarkable finding was that, for low-to-moderate viscosity ratios, the axisymmetry of the base flow may break down through an \emph{internal} flow instability. That is, the flow asymmetry is initially generated and amplified solely within the droplet. This mechanism differs fundamentally from the \emph{external} flow instability typically encountered in flow past bubbles \citep{2007_Magnaudet, yang2007linear, tchoufag2013linear, legendre2025gas}, particles \citep{fabre2008bifurcations, tchoufag2014global}, and highly viscous droplets ($\mu^\ast\gg1$) \citep{shi2025wake}, for which the internal flow plays no significant role and the instability first develops in the near wake. In the internal-instability scenario, the wake axisymmetry also eventually breaks down, but only at a later stage. The wake then evolves towards a sequence of new equilibrium states, including, in particular, biplanar-symmetric and uniplanar-symmetric states, as well as a bistable state depending on the initial conditions. In the present work, we consider the free rise of a buoyant droplet in an immiscible liquid for drop-to-fluid viscosity ratios $\mu^\ast=\mathcal{O}(0.1-1)$. We focus on the regime in which flow axisymmetry breaks down through the internal instability described above, and examine how the droplet dynamics responds to the various non-axisymmetric wake patterns selected by the coupled fluid--droplet system.

As reviewed in \citet{2012_Ern} and \citet{2025_Shi_drop, shi2025wake}, previous studies of the path instability of three-dimensional axisymmetric bodies have focused mainly on bubbles and particles. For such bodies, the primary flow instability observed when the body is held fixed is closely connected with the onset of the first non-vertical path when the same body is free to move. For instance, in the case of a high-Reynolds-number clean (i.e. free of surfactants) bubble, the threshold for flow instability corresponds to a critical major-to-minor axis ratio of about 2.1 \citep{yang2007linear, 2007_Magnaudet, tchoufag2013linear}, close to the corresponding threshold (about 2.0) for the first path instability of a freely rising bubble \citep{1995_Duineveld, 2008_Zenit, 2016_Cano-Lozano, 2024_Bonnefis, 2025_Shi, legendre2025gas}. Similarly, for a spherical particle, the first flow instability sets in at a Reynolds number close to 210 \citep{ghidersa2000breaking, fabre2008bifurcations}, in good agreement with the range $\Rey \in [210,260]$ within which the first path instability of a freely rising light sphere is observed \citep{2004_Jenny, 2010_Horowitz, 2018_Auguste}.

By comparison, investigations of the path instability of freely rising or settling droplets remain rather limited. To the best of our knowledge, the only directly relevant study is that of \citet{2015_Albert}, who carried out direct numerical simulations (DNS) of corn-oil droplets rising in pure water, for which $\mu^\ast\approx46$, over a wide range of droplet radii $R$. As $R$ increases, the sequence of non-vertical paths follows approximately that reported for light spherical particles (see, e.g. \citealp{2018_Auguste}) up to the critical size at which shape oscillations set in, beyond which the path becomes fully three-dimensional and chaotic. An interesting point is that the first non-vertical path is a steady oblique path and occurs at a critical Reynolds number close to the threshold of the external flow instability of a spherical particle. Most other investigations of freely rising or settling droplets have instead focused on the variation of the terminal rise speed with droplet size, their main objective being to propose empirical drag correlations for use as closure relations in industrial-scale simulations of droplet-laden suspensions; see \citet{loth2008quasi} for a review of early work, and \citet{abdelouahab2011new}, \citet{zhang2019state} and \citet{2026_gode} for overviews of more recent studies. Among these systems, toluene droplets rising in water, for which $\mu^\ast=0.62$, have received particular attention. The corresponding line of work stems from the detailed experiments of \citet{wegener2009einfluss}, who identified at least three critical droplet sizes at which the trend followed by the terminal rise speed changes abruptly. Specifically, beyond a first critical size $R_{c1}\approx1.1\,\text{mm}$, the rise speed starts to decrease with increasing $R$; as the droplet size exceeds approximately $1.5\,\text{mm}$, it suddenly increases by more than $50\%$; finally, beyond approximately $2.3\,\text{mm}$, the rise speed starts to oscillate in time.

To interpret the origin of these characteristic sizes, it is natural to seek a connection with instabilities of the flow past the same body held fixed. For the present system, however, this connection is not straightforward. Consider, for instance, the decrease in the rise speed beyond the first critical size. Although the first external flow instability \citep{2007_Magnaudet} is known to enhance the drag and hence reduce the terminal velocity (see, e.g. \citealp{veldhuis2009freely, 2010_Horowitz, 2018_Auguste}) when the body is free to move, the low viscosity ratio $\mu^\ast=0.62$ and the fact that the droplet remains nearly spherical at this stage make the onset of such an external instability very unlikely; see \citet{2025_Shi_drop} for a quantitative justification.

Since the detailed experiments of \citet{wegener2009einfluss}, several DNS studies have attempted to reproduce the first critical size of toluene droplets rising in water. Early simulations in this direction \citep{2011_Baumler, engberg2014numerical, wegener2014numerical} were carried out in an axisymmetric configuration, based on the expectation that the underlying drag enhancement might result from repeatable shape oscillations \citep{magnaudet2011reciprocal, lalanne2013effect, 2025_Shi}. These simulations failed to predict the above value of $R_{c1}$: the droplet shape remained stable until $R$ exceeded about $2.1\,\text{mm}$ \citep{2011_Baumler}, i.e. almost twice $R_{c1}$. This discrepancy motivated subsequent fully resolved three-dimensional DNS \citep{bertakis2010validated, eiswirth2011experimental}. Owing to the slow development of the axisymmetry-breaking process, and hence to the long physical time required, a reasonable reproduction of the first two characteristic sizes was achieved only in the more recent studies of \citet{2015_Engberg} and \citet{charin2019dynamic}. In particular, these authors showed that, near $R\approx1.5\,\text{mm}$, two distinct terminal rise speeds may be obtained by changing the initial shape and orientation of the droplet, in line with the experimental observations of \citet{wegener2009einfluss}. In both studies, the largest droplet radius considered was $2.0\,\text{mm}$, up to which the terminal rise speed remained steady.

A parallel line of investigation has explored the possible existence of a distinct flow-destabilization mechanism specific to liquid--liquid systems with low-to-moderate viscosity ratios. The first step in this direction was taken by \citet{edelmann2017numerical}, who reported three-dimensional DNS of a uniform flow past a spherical droplet at Reynolds numbers of $\mathcal{O}(100)$. Remarkably, their results showed that, at a viscosity ratio of 0.5, i.e. close to that of toluene in water, the flow axisymmetry may break down, with two pairs of streamwise vortices emerging in the wake and producing a marked drag increase. This form of axisymmetry breaking differs fundamentally from the external flow instability described by \citet{2007_Magnaudet}: in the latter case, the bifurcation is driven by an azimuthal mode of wavenumber $m=1$, so that only one pair of streamwise vortices is present in the wake after the bifurcation. Subsequent and more systematic numerical investigations \citep{rachih2019etude, gode2024flow, 2024_Shi_drop, 2025_Zhang, 2026_gode} considered different ranges of governing parameters, including the viscosity ratio and the external and internal Reynolds numbers (definitions to be given in the next section). These studies confirmed the flow-instability scenario reported by \citet{edelmann2017numerical} and showed that, for given internal and external Reynolds numbers, this instability persists below a critical viscosity ratio ranging from $\mathcal{O}(0.1)$ to 10. 

The fixed-droplet problem was further examined in our recent study \citep{2025_Shi_drop}. By monitoring separately the evolution of the azimuthal perturbation energy inside and outside the droplet, we showed that the axisymmetry breaking associated with this instability first develops solely within the droplet. This instability is therefore referred to hereafter as an internal flow instability. Drawing an analogy with the argument proposed by \citet{2007_Magnaudet} for the external flow instability, and building on the DNS results available so far, we proposed a criterion for its onset based on the maximum internal vorticity in the corresponding base flow.

Combining the knowledge gained from the fixed-droplet configuration \citep{2025_Shi_drop} with that obtained for freely rising toluene droplets in water \citep{wegener2009einfluss, 2011_Baumler} raises several open issues. First, the criterion for the onset of the first internal flow instability predicts a critical radius $R_{c1}\approx0.97\,\text{mm}$ for a toluene droplet rising in water (see figure 19 of \citealp{2025_Shi_drop}), slightly below the experimental value $R_{c1}\approx1.1\,\text{mm}$. Second, the existence, in the fixed-droplet configuration, of a parameter range in which biplanar- and uniplanar-symmetric flow structures coexist suggests that, in the free-rise problem, vertical and oblique paths might similarly coexist. Such a bistable rise regime has not been reported so far. The closest related observation concerns the case $R\approx1.5\,\text{mm}$, for which two distinct terminal rise speeds have been identified \citep{wegener2009einfluss, 2015_Engberg, charin2019dynamic}. Finally, in the experiments, both the rise speed and the droplet shape exhibit sustained oscillations when $R$ exceeds approximately $2.3\,\text{mm}$. No such unsteady regime was observed in the fixed-droplet simulations of \citet{2025_Shi_drop}, in which the droplet was prescribed to be spherical and hence non-deformable. Shape oscillations were indeed reported in the axisymmetric DNS of freely rising deformable droplets by \citet{2011_Baumler}, but the critical size beyond which they set in was about $2.1\,\text{mm}$, smaller than that found experimentally.

The present work aims to address these open issues, and more generally to provide systematic DNS results for the dynamics of freely rising droplets in a liquid--liquid system of low-to-moderate viscosity ratio. Such results have not been available so far, and provide original reference data for future comparisons with experiments, DNS and global linear-stability analyses of the path instability of freely moving droplets. The paper is organised as follows. In \S\,\ref{sec:problem_state}, we formulate the problem and outline the numerical approach. Section~\ref{sec:res_2d} discusses results obtained by constraining the flow to remain axisymmetric, while fully three-dimensional evolutions are examined in \S~\ref{sec:res_3d}. The influence of the initial conditions on the rise dynamics is discussed in \S~\ref{sec:inf_dis}. The main findings and remaining open issues are summarised in \S~\ref{sec:conclusion}.

\section{Problem statement and numerical approach}
\label{sec:problem_state}
We consider a single droplet of volume $\mathcal{V}$ and equivalent radius $R=(3\mathcal{V}/4\pi)^{1/3}$ rising freely in a large body of an immiscible Newtonian liquid, under the combined effects of gravity $g$ and interfacial tension $\gamma$. The density and dynamic viscosity of the droplet are denoted by $\rho^i$ and $\mu^i$, respectively, while those of the surrounding liquid are denoted by $\rho^e$ and $\mu^e$. We focus on liquid--liquid systems with $\mu^i/\mu^e=\mathcal{O}(1)$ and $\rho^i/\rho^e<1$. The droplet therefore has a viscosity comparable to that of the external liquid and rises owing to buoyancy. The external liquid is at rest at infinity and the flow induced by the droplet ascent is assumed to be incompressible. The instantaneous velocity and position of the droplet centroid are denoted by $\boldsymbol{v}$ and $\boldsymbol{x}$, respectively. The liquid velocity in a frame of reference translating with the droplet is denoted by $\boldsymbol{u}$, and the corresponding vorticity is $\boldsymbol{\omega}=\boldsymbol{\nabla}\times\boldsymbol{u}$. Unless stated otherwise, the droplet is initially spherical and is released from rest.

The problem generally depends on four dimensionless parameters, namely the density ratio $\rho^\ast$, the viscosity ratio $\mu^\ast$, the Bond number $Bo$ and the Galilei number $Ga$, defined as
\begin{equation}
\rho^\ast=\frac{\rho^i}{\rho^e},\qquad \mu^\ast=\frac{\mu^i}{\mu^e},\qquad
Bo=\frac{\rho^e g |1-\rho^\ast|R^2}{\gamma},\qquad
Ga=\frac{\rho^e R \left(|1-\rho^\ast|gR\right)^{1/2}}{\mu^e}.
\label{eq:dimensionless_parameters}
\end{equation}
The last two parameters compare the buoyancy force driving the droplet ascent, of order $\rho^e |1-\rho^\ast|gR^3$, with the capillary force, of order $\gamma R$, and the viscous force, of order $\mu^e (gR^3)^{1/2}$, respectively. Once the instantaneous droplet velocity $\boldsymbol{v}(t)$ is known, one may also define the instantaneous external and internal Reynolds numbers as
\begin{equation}
Re^e=\frac{2\rho^e R ||\boldsymbol{v}||}{\mu^e},\qquad
Re^i=\frac{2\rho^i R ||\boldsymbol{v}||}{\mu^i}.
\end{equation}

For a given liquid--liquid system in a prescribed gravitational environment, the material properties may be gathered into the Morton number,
\begin{equation}
Mo=\frac{Bo^3}{Ga^4} =\frac{g(\mu^e)^4 |1-\rho^\ast|}{\rho^e\gamma^3}.
\end{equation}
This quantity is independent of the droplet size and therefore remains fixed when $R$ is varied at constant fluid properties.

To establish a direct connection with previous studies, we focus on toluene droplets rising in water, for which $\rho^\ast=0.865$, $\mu^\ast=0.62$ and $Mo=1.95\times10^{-11}$. The corresponding physical properties are listed in table~\ref{tab:phy_par}. With this choice, the droplet radius $R$ is the only remaining control parameter, and will be used throughout to label the different cases. We consider radii in the range $0.5\,\text{mm}\leq R\leq 3\,\text{mm}$, corresponding approximately to $0.01\leq Bo\leq0.34$ and $15\leq Ga\leq210$. Over this range, the droplet shape evolves from nearly spherical at small $R$ to increasingly oblate, approximately ellipsoidal shapes at larger $R$. This deformation may be characterised by the aspect ratio $\upchi=b/a$, where $b$ and $a$ denote the major and minor axes, respectively. In the imposed axisymmetric configuration, $\upchi$ provides a natural measure of the deformation level. In fully three-dimensional regimes, especially when the droplet shape is no longer axisymmetric, a single aspect ratio is no longer sufficient; we therefore also use the dimensionless interfacial area $\Sigma=A/(4\pi R^2)$, where $A$ is the instantaneous liquid--liquid interfacial area, as a scalar measure of the overall deformation level. The terminal-state results discussed below indicate that, as $R$ increases from $0.5$ to $3\,\text{mm}$, $\upchi$ increases from nearly unity to about $2.5$, while both $Re^e$ and $Re^i$ increase from $\mathcal{O}(100)$ to approximately $1000$.

\begin{table}
\centering
\begin{tabular}{lllllll}
%\hline
Liquid & $\rho~(\text{kg~m}^{-3})~~$ & $\mu~(\text{mPa~s})~~$ & $\gamma~(\text{mN~m}^{-1})~~$ & $\rho^{*}$ & $\mu^{*}$ & $Mo$ \\
%\hline
Toluene~~ & 862.3  & 0.55  & 35 & 0.865~~ & 0.62~~ & $1.95\times 10^{-11}$ \\
Water   & 997.0 & 0.89 &    &       &      &  \\
%\hline
\end{tabular}
\caption{Physical parameters for toluene and water at $T = 25^\circ$C.}
\label{tab:phy_par}
\end{table}

The results discussed below were obtained by solving the time-dependent two-phase Navier--Stokes equations, either in an axisymmetric or in a fully three-dimensional configuration, using the open-source flow solver Basilisk \citep{2009_Popinet, 2015_Popinet}. The flow inside and outside the droplet is computed within a one-fluid formulation, assuming that the interface is free of contamination. The interface evolution is described by the transport equation for the volume fraction $C(\boldsymbol{x},t)$, with $C=1$ inside the droplet and $C=0$ in the external liquid. The local density and dynamic viscosity are evaluated from $C$ using, respectively, arithmetic and harmonic averaging. It should be kept in mind that the harmonic averaging of the dynamic viscosity does not guarantee a continuous variation of the shear stresses across the interface \citep{kothe1998perspective, magnaudet2025rational}; a sufficiently fine resolution is therefore required to make the associated numerical artefact negligible. Surface tension is modelled using the balanced-force formulation of \citet{francois2006balanced}, which is based on the continuum-surface-force approach of \citet{brackbill1992continuum}. The curvature entering the capillary force is computed with the height-function technique of \citet{2009_Popinet}. Further details on the numerical schemes implemented in Basilisk may be found in \citet{popinet2018numerical}.

In the three-dimensional simulations, the computational domain is a cubic box of edge length $L=960R$, large enough to follow the droplet over a sufficiently long vertical distance and to examine its behaviour once a fully developed state has been reached. No-slip and no-penetration conditions are imposed on all domain boundaries. The droplet is initially released at a vertical distance $15R$ above the bottom wall, so as to minimize any confinement effect induced by this boundary during the early stages of the rise. The domain is discretized using the adaptive mesh-refinement (AMR) technique implemented in Basilisk \citep{2018_Hooft}. The refinement criteria are selected following those used in our previous studies \citep{2024_Shi_bub, 2025_Shi}, such that the minimum cell size satisfies $\Delta_{\min}/R\approx1/68$ in the vicinity of the interface and $\Delta_{\min}/R\approx1/34$ in the far wake, located typically more than $10R$ downstream of the droplet centroid. With this resolution, at least three grid cells lie within the boundary layers developing on both sides of the interface for Reynolds numbers up to approximately $1000$. Further details on the AMR criteria, the grid structure near the droplet and in the far wake, as well as on the computational resources and typical wall-clock times required by the simulations, may be found in the two studies referenced above.

The numerical approach, together with the grid resolution resulting from the AMR criteria just described, has already been shown to reliably predict the various lateral motions of a bubble rising near a vertical wall at Reynolds numbers, both $\Rey^e$ and $\Rey^i$, up to approximately $1000$ \citep{2024_Shi_bub,2025_Shi}. To further assess its suitability for freely rising droplets, we carried out in \S~\ref{sec:res_2d} a grid-convergence study in an axisymmetric configuration. Reducing the minimum cell size from approximately $R/68$ to $R/136$ was found to produce negligible changes in both the terminal rise speed and the droplet shape over the entire range $0.5\,\text{mm}\leq R\leq3.0\,\text{mm}$. For $R\gtrsim2.05\,\text{mm}$, the same axisymmetric DNS also predict the onset of repeatable shape oscillations which are in quantitative agreement with previous simulations \citep{2011_Baumler,engberg2014numerical}. In \S~\ref{sec:res_3d}, the terminal rise speeds obtained in the fully three-dimensional simulations are further compared with the experimental data of \citet{wegener2009einfluss} and the numerical results of \citet{2015_Engberg} and \citet{charin2019dynamic}. The reasonable agreement obtained in these comparisons confirms the reliability of the present numerical approach for the problem under consideration.

Unless stated otherwise, all quantities reported below are made dimensionless using $R$ and $\sqrt{R/g}$ as characteristic length and time scales, respectively. The dimensionless time is denoted by $T$, while the dimensionless position and velocity of the droplet centroid are denoted by $\boldsymbol{X}$ and $\boldsymbol{V}$, respectively. A dimensional frequency $f$ is generally reported below in dimensionless form as $\overline{f}=f\sqrt{R/g}$. When comparison with previous studies of path oscillations or vortex shedding is useful, we also use the Strouhal number $St=2fR/v_m$, where $v_m$ denotes the dimensional mean rise speed. The vorticity of the disturbance flow is normalised by $\sqrt{g/R}$ and denoted by $\overline{\boldsymbol{\omega}}$. When needed, either a cylindrical coordinate system $(r,\theta,y)$ or a Cartesian coordinate system $(x,y,z)$ will be introduced, with its precise definition specified in the relevant section. The component of a vector along the $i$ coordinate direction is denoted by the subscript $i$.

\section{Axisymmetric configuration} \label{sec:res_2d}

An axisymmetric configuration is first considered to characterize the droplet shape, rise velocity and flow structure associated with the base state. For this purpose, instead of the cubic domain described above, we use an axisymmetric domain of radius $480R$ and height $960R$. To ease the discussion, we introduce a cylindrical coordinate system $(r,\theta,y)$ whose axis $r=0$ coincides with the symmetry axis. The origin of the vertical coordinate is chosen such that the droplet centroid is initially located at $X_y=0$. The droplet is released on the symmetry axis and rises along the $+y$ direction.

\begin{figure}
\centerline{\includegraphics[scale=0.6]{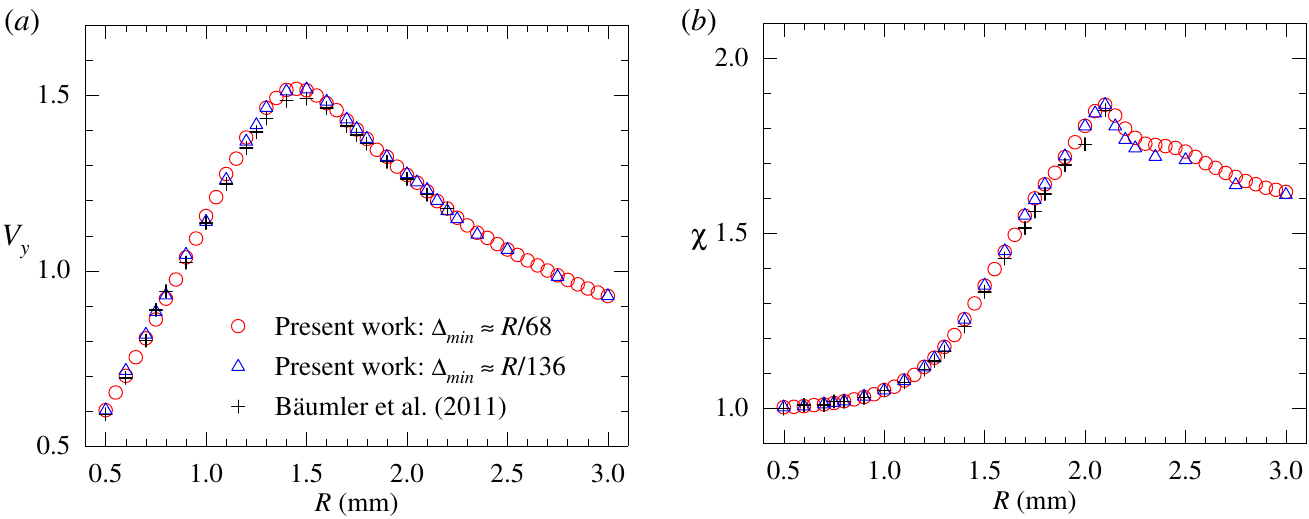}}
\caption{Fully developed state in the imposed axisymmetric configuration. $(a)$ Rise velocity $V_y$; $(b)$ aspect ratio $\upchi$. For $R\gtrsim2.1\,\text{mm}$, where periodic oscillations occur, the reported values are averaged over one oscillation period.}
\label{fig:v_vs_R_2d}
\end{figure}

Figure~\ref{fig:v_vs_R_2d} shows the terminal rise velocity $V_y$ (averaged over one period when oscillations are observed) and aspect ratio $\upchi$ of the droplet in the fully developed state for $R$ increasing from $0.5$ to $3.0\,\text{mm}$. Taking advantage of the reduced computational cost of the axisymmetric configuration, two series of simulations were carried out to assess grid convergence, with minimum cell sizes $\Delta_{\min}=R/68$ and $\Delta_{\min}=R/136$, respectively. For both $V_y$ and $\upchi$, the predictions obtained with the two spatial resolutions are in very good agreement, confirming that $\Delta_{\min}=R/68$ is sufficient over the entire range of $R$ considered. Unless stated otherwise, the axisymmetric results of the present work discussed below were obtained with $\Delta_{\min}=R/68$.

Also shown in figure~\ref{fig:v_vs_R_2d} are the axisymmetric predictions of \citet{2011_Baumler}, obtained with a boundary-fitted finite-element moving-mesh method in a reference frame translating with the droplet. This numerical approach is quite different from the present VOF/AMR method. The data of \citet{2011_Baumler}, available up to $R=2.1\,\text{mm}$, agree very well with the present predictions for both $V_y$ and $\upchi$. This agreement provides an additional cross-method check of the present axisymmetric results. We also note that \citet{engberg2014numerical} later considered the same axisymmetric toluene--water problem using a level-set finite-volume method on a structured grid refined around the droplet, and reported aspect ratios almost identical to those of \citet{2011_Baumler}.

\begin{figure}
\centerline{\includegraphics[scale=0.75]{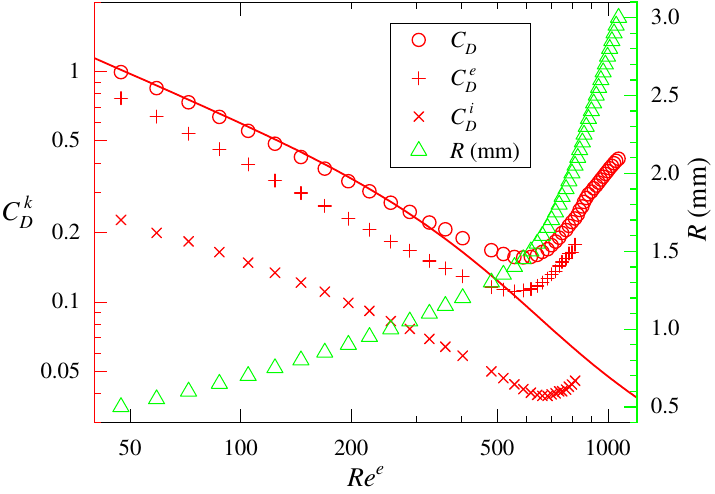}}
\caption{Drag coefficient and droplet radius as functions of the external Reynolds number $\Rey^e$. Red circles, left axis: total drag coefficient $C_D$ obtained from the balance between drag and buoyancy using Eq.~\eqref{eq:cd_vs_v}; green symbols, right axis: droplet radius $R$ corresponding to each value of $\Rey^e$; red solid line: prediction of the drag model of \citet[see Eqs.~(3.3)--(3.4) therein]{2024_Shi_drop}. The drag contributions $C_D^i$ and $C_D^e$, associated with viscous dissipation inside and outside the droplet, respectively, are calculated using Eq.~\eqref{eq:cd_vs_disp2}.}
\label{fig:cd_vs_re_2d}
\end{figure}

The variation of $V_y$ with $R$ shown in figure~\ref{fig:v_vs_R_2d} reveals some interesting features regarding the dependence of the drag coefficient on the external Reynolds number $\Rey^e$. In the terminal state, the balance between drag and buoyancy gives
\begin{equation}
\frac{1}{2} \pi R^2 \rho^e C_D\, (V_y\sqrt{gR})^2 = \rho^e |1-\rho^\ast| \frac{4}{3} \pi R^3 g,
\label{eq:drag_vs_buoy}
\end{equation}
where $C_D$ is the drag coefficient. Rearranging Eq.~\eqref{eq:drag_vs_buoy}, one obtains
\begin{equation}
C_D =\frac{8}{3}|1-\rho^\ast|V_y^{-2}.
\label{eq:cd_vs_v}
\end{equation}
On the other hand, the external Reynolds number is related to the Galilei number $Ga$ and the dimensionless rise velocity $V_y$ through
\begin{equation}
\Rey^e = 2\,Ga\,V_y\,|1-\rho^\ast|^{-1/2}.
\label{eq:re_vs_v}
\end{equation}
Thus, once the dimensionless rise velocity $V_y$ is known for a given $R$, and hence for a given $Ga$, the corresponding values of $C_D$ and $\Rey^e$ follow directly from Eqs.~\eqref{eq:cd_vs_v} and \eqref{eq:re_vs_v}. Figure~\ref{fig:cd_vs_re_2d} shows the resulting drag coefficient as a function of $\Rey^e$; the corresponding droplet radius is also indicated. Remarkably, the results reveal the existence of a critical external Reynolds number $\Rey^e_c\approx615$, corresponding to $R\approx1.5\,\text{mm}$. For $\Rey^e<\Rey^e_c$, $C_D$ decreases with increasing $\Rey^e$, whereas the opposite trend is observed for $\Rey^e>\Rey^e_c$. Also shown in figure~\ref{fig:cd_vs_re_2d} is the prediction of the drag model of \citet[see Eqs.~(3.3)--(3.4) therein]{2024_Shi_drop}, which is expected to apply to nearly spherical droplets up to $\Rey^e\approx1000$. This prediction agrees well with the present results up to $\Rey^e\approx325$, beyond which it starts to underestimate the drag. This underestimation can be understood by noting that the droplet aspect ratio already reaches $\upchi\approx1.1$ at $R=1.1\,\text{mm}$, corresponding to $\Rey^e\approx325$; see figure~\ref{fig:v_vs_R_2d}$(b)$. Larger droplets adopt a nearly oblate shape, so that their frontal area in the cross-stream plane, and hence their drag coefficient, increases with $\upchi$ and is larger than that of a spherical droplet at the same $\Rey^e$.

\begin{figure}
\centerline{\includegraphics[scale=0.6]{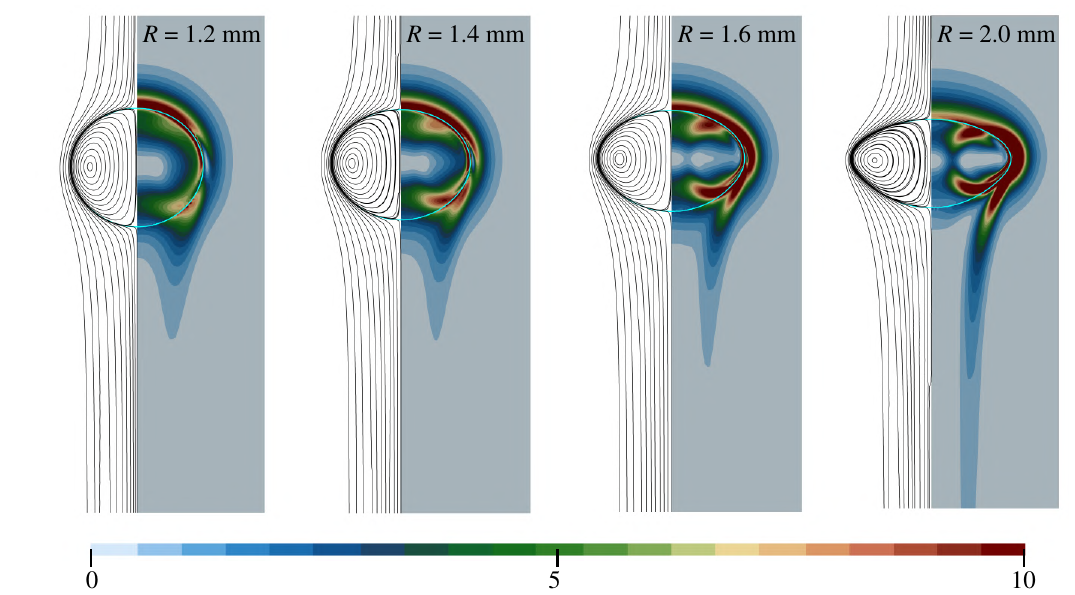}}
\caption{Base flow past droplets of different radius $R$. In each panel, the droplet rises upwards so that the relative flow is from top to bottom. The left half shows streamlines in the frame translating with the droplet, while the right half shows contours of the normalised viscous dissipation rate $\overline{\Phi}$.}
\label{fig:streamline_disp}
\end{figure}

For $\Rey^e\gtrsim325$, corresponding to $R\gtrsim1.1\,\text{mm}$, two competing effects therefore influence the drag coefficient. The first is associated with droplet deformation : as the droplet becomes more oblate, its actual frontal area increases, so that the corresponding increase in drag is reflected here as an increase in $C_D$, since $C_D$ is normalised using the fixed reference area $\pi R^2$. The second is associated with the potential-flow mechanism characteristic of low-to-moderate-viscosity-ratio droplets moving at $\Rey^e=\mathcal{O}(100)$: at fixed $\upchi$, this mechanism tends to make $C_D$ decrease as $(\Rey^e)^{-1}$ for sufficiently large $\Rey^e$ \citep{moore1965velocity,stone1993interpretation,blanco1995structure}. The fact that $C_D$ starts to increase with $\Rey^e$ for $\Rey^e\gtrsim615$ indicates that the former effect then overcomes the latter. A plausible explanation for this drag increase could be the formation of a standing eddy in the near wake of the droplet, which is known to enhance the pressure drag \citep{dandy1989buoyancy,magnaudet1995accelerated}. To check this possibility, figure~\ref{fig:streamline_disp} shows, in the left half of each panel, the streamlines obtained in the terminal state in a frame translating with the droplet. Although the streamlines become increasingly distorted near the rear of the droplet as $R$ increases, no flow separation is observed, even up to $R=2.0\,\text{mm}$.

Another useful perspective is provided by the energy budget. In the terminal state, the rate of work done by the drag balances the viscous dissipation integrated over the whole fluid domain, namely
\begin{equation}
\frac{1}{2} \pi R^2 \rho^e C_D\, ||\boldsymbol{v}||^3 = \int_{\mathcal{V}} \Phi ~\mathrm{d}\mathcal{V},
\label{eq:cd_vs_disp1}
\end{equation}
where $\Phi=2\mu(\boldsymbol{x})\mathcal{S}:\mathcal{S}$ is the viscous dissipation rate per unit volume, with $\mathcal{S}=\frac{1}{2}\left(\boldsymbol{\nabla}\boldsymbol{u}+(\boldsymbol{\nabla}\boldsymbol{u})^T\right)$ denoting the strain-rate tensor. Rearranging Eq.~\eqref{eq:cd_vs_disp1} gives
\begin{equation}
C_D = \frac{4}{\pi}\int_{\overline{\mathcal{V}}} \frac{1}{\Rey^e}\overline{\Phi} ~\mathrm{d}\overline{\mathcal{V}} = \underbrace{\frac{4}{\pi}\int_{\overline{\mathcal{V}}^i} \frac{1}{\Rey^e}\overline{\Phi} ~\mathrm{d}\overline{\mathcal{V}}^i}_{C_D^i} + \underbrace{\frac{4}{\pi}\int_{\overline{\mathcal{V}}^e} \frac{1}{\Rey^e}\overline{\Phi} ~\mathrm{d}\overline{\mathcal{V}}^e}_{C_D^e},
\label{eq:cd_vs_disp2}
\end{equation}
where $\overline{\Phi}=\Phi/(\mu^e R||\boldsymbol{v}||^2)$ is the normalised dissipation rate and $\overline{\mathcal{V}}^k=\mathcal{V}^k/R^3$ is the dimensionless volume of either the internal ($k=i$) or external ($k=e$) fluid domain. Equation~\eqref{eq:cd_vs_disp2} therefore decomposes the drag coefficient into two contributions, $C_D^i$ and $C_D^e$, associated with viscous dissipation inside and outside the droplet, respectively.

The two drag contributions calculated using Eq.~\eqref{eq:cd_vs_disp2} are also shown in figure~\ref{fig:cd_vs_re_2d}. For all $R$, the difference between the drag coefficient obtained from the force balance, Eq.~\eqref{eq:cd_vs_v}, and the sum $C_D^i+C_D^e$ remains below $3\%$, confirming the consistency of the two estimates. Moreover, $C_D^i$ remains smaller than $30\%$ of $C_D$, especially for $R\geq1.5\,\text{mm}$, indicating that the increase in $C_D$ observed for $\Rey^e\gtrsim615$ is mainly associated with enhanced dissipation in the external fluid. This is confirmed by the contours of the normalised viscous dissipation rate $\overline{\Phi}$ shown in the right half of each panel of figure~\ref{fig:streamline_disp}. For all $R$, the most intense dissipation is located within the external boundary layer, consistent with the relatively small value of $C_D^i/C_D$ discussed above. The key difference between droplets with $R\lesssim1.5\,\text{mm}$ and larger droplets lies in the location of the region where dissipation is significant. For $R\lesssim1.5\,\text{mm}$, this region remains confined to the front part of the droplet, whereas for larger $R$ it shifts towards the equatorial region, where the surface curvature reaches its maximum. This topological change, namely the pronounced dissipation developing near the equatorial region for $R\gtrsim1.5\,\text{mm}$, appears to be the main cause of the increase in $C_D$ for $\Rey^e\gtrsim615$.

\begin{figure}
\centerline{\includegraphics[scale=0.34]{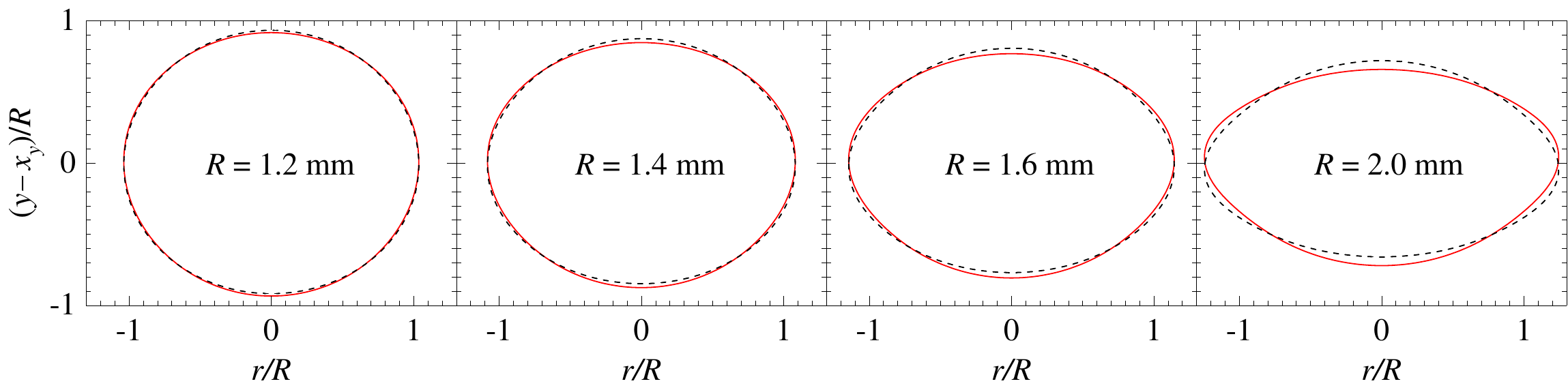}}
\caption{Terminal shapes of droplets of different radius $R$. The red solid line denotes the actual droplet contour, while the black dashed line shows the same contour reflected with respect to the centroidal horizontal line $y=X_y$, in order to highlight the fore--aft asymmetry of the droplet shape.}
\label{fig:vof_2d}
\end{figure}

Figure~\ref{fig:vof_2d} shows the terminal droplet shapes for selected values of $R$. In each panel, the red solid line represents the actual droplet contour, while the black dashed line is obtained by reflecting this contour with respect to the horizontal line passing through the droplet centroid, $y=X_y$. The comparison between these two contours shows that the droplet remains nearly fore--aft symmetric up to $R\approx2.0\,\text{mm}$, at which point a visible asymmetry starts to develop.

\begin{figure}
\centerline{\includegraphics[scale=0.6]{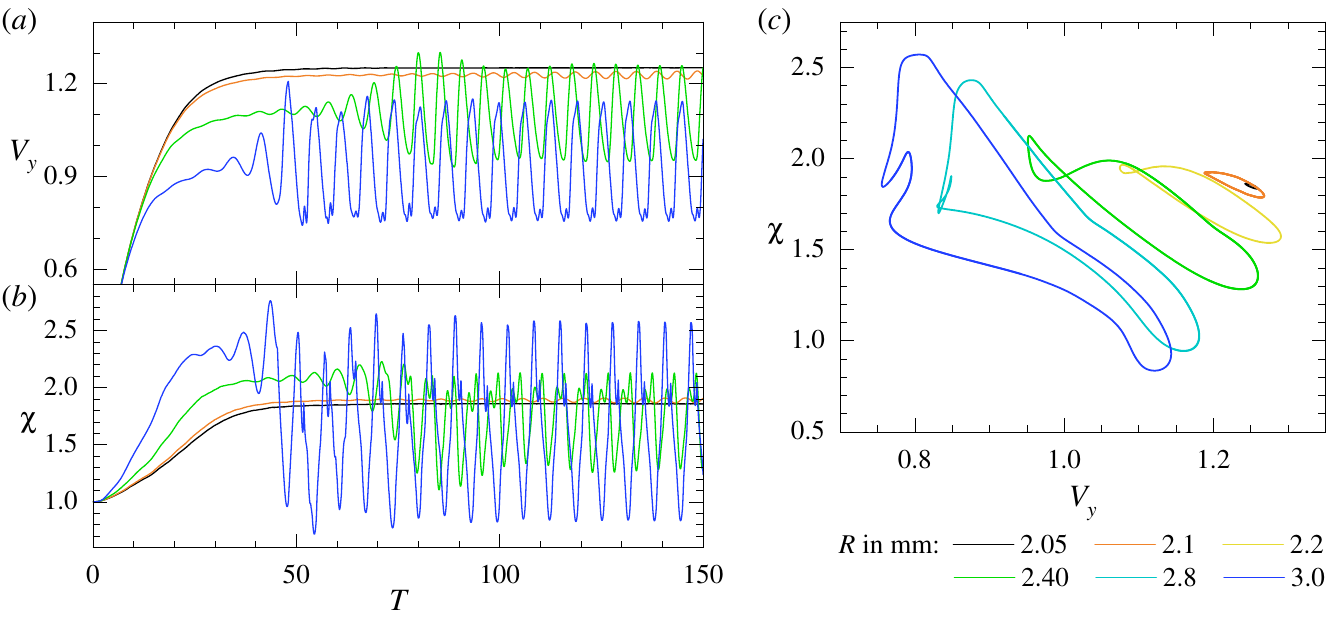}}
\caption{Time evolution and phase relation of droplets with $R\geq2.05\,\text{mm}$ in the imposed axisymmetric configuration. $(a)$ Rise speed $V_y$; $(b)$ aspect ratio $\upchi$; $(c)$ variation of $V_y(t)$ as a function of $\upchi(t)$ over one oscillation period in the fully developed stage. In $(c)$, time increases in the anticlockwise direction; in most cases, $V_y$ increases as $\upchi$, and hence the frontal area in the cross-stream plane, decreases, and vice versa.}
\label{fig:v_chi_large_R}
\end{figure}

The terminal state becomes unsteady for $R\gtrsim2.05\,\text{mm}$. Figure~\ref{fig:v_chi_large_R}$(a)$ shows the time evolution of the rise speed $V_y$ for $R$ increasing from $2.05$ to $3.0\,\text{mm}$. During the long-time evolution, say for $T>100$, $V_y$ reaches a nearly constant value for $R=2.05\,\text{mm}$, whereas it undergoes regular oscillations for $R\geq2.1\,\text{mm}$. The relative amplitude of these oscillations increases with $R$, from approximately $3.7\%$ at $R=2.1\,\text{mm}$ to about $20\%$ at $R=3.0\,\text{mm}$. Irrespective of the droplet size, the Strouhal number associated with these primary oscillations remains close to $0.33$. Although not shown here, the mean rise speed, oscillation amplitude and frequency obtained for $R=2.2\,\text{mm}$ agree well with those reported by \citet{engberg2014numerical}.

Figure~\ref{fig:v_chi_large_R}$(b)$ shows the corresponding evolution of the aspect ratio $\upchi$. For $R\geq2.1\,\text{mm}$, $\upchi$ oscillates at the same frequency as $V_y$. The phase relation between these two quantities is displayed more clearly in figure~\ref{fig:v_chi_large_R}$(c)$, which shows the variation of $V_y(t)$ as a function of $\upchi(t)$ over one oscillation period in the fully developed stage. Along each curve, time increases in the anticlockwise direction. The rise speed reaches a local maximum shortly after $\upchi$ reaches a local minimum, and the same correspondence holds between the local minima of $V_y$ and the local maxima of $\upchi$. This phase relation indicates that the oscillations of $V_y$ are primarily driven by droplet-shape oscillations, rather than by an instability of the disturbance flow field. Indeed, in the latter case, the deformation level is expected to respond to the inertia of the flow, and hence to the rise speed, so that $\upchi$ would reach a local maximum shortly after $V_y$ reaches a maximum, with a similar correspondence between their local minima \citep{2025_Shi_drop,2025_Shi}.

\section{Three-dimensional configuration} \label{sec:res_3d}

\subsection{Overview of rise regimes} \label{sec:overview}
As mentioned above, previous studies \citep{wegener2009einfluss,wegener2010terminal,2015_Engberg,charin2019dynamic} have shown that, as the droplet size $R$ increases beyond about $1.1\,\text{mm}$, the flow axisymmetry breaks down and the terminal rise speed $V_y$ decreases with increasing $R$. They also reported the existence of a second characteristic size, about $1.5\,\text{mm}$, beyond which the flow axisymmetry may recover. We performed fully three-dimensional time-dependent DNS for $R$ up to $3.0\,\text{mm}$ to examine these transitions and their consequences for both the rise speed and the droplet path.

\begin{figure}
\centerline{\includegraphics[scale=0.65]{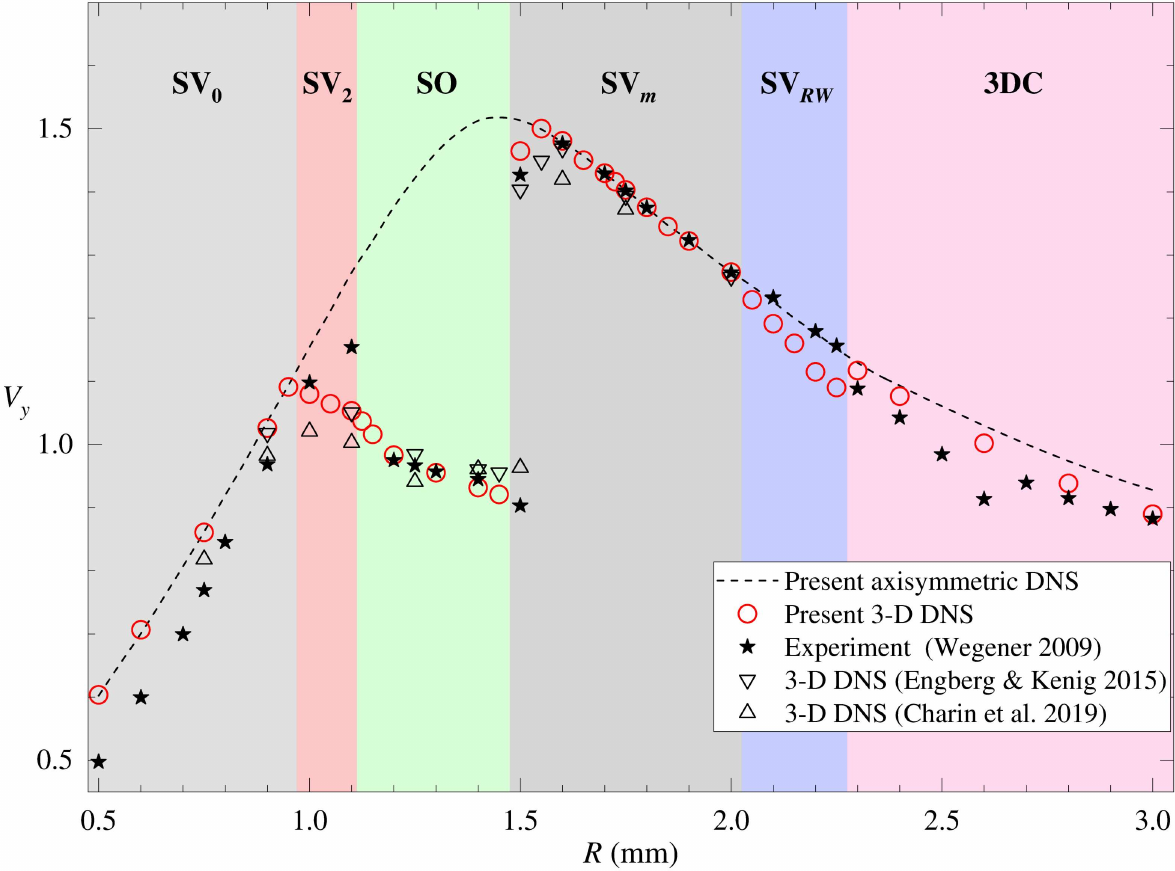}}
\caption{Terminal rise speed of the droplet as a function of the droplet radius $R$. The black dashed line and red open circles denote the present axisymmetric and fully three-dimensional DNS, respectively. Black stars correspond to the experimental data of \citet{wegener2009einfluss}, while the black inverted and upright triangles denote the three-dimensional DNS results of \citet{2015_Engberg} and \citet{charin2019dynamic}, respectively. As $R$ increases, six regimes are encountered: the steady vertical regime ($\text{SV}_0$), in which the flow remains axisymmetric; the secondary steady vertical regime ($\text{SV}_2$), in which the flow is biplanar-symmetric; the steady oblique regime (SO), in which the flow is uniplanar-symmetric; the third steady vertical regime ($\text{SV}_m$), in which the flow is either axisymmetric or only weakly asymmetric; the rotating-wave regime ($\text{SV}_{RW}$), in which the droplet rises along a nearly vertical path while undergoing an approximately steady axial spinning motion; and finally the three-dimensional chaotic regime (3DC), in which the path is fully three-dimensional and chaotic.}
\label{fig:v_vs_R_3d}
\end{figure}

\begin{figure}
\centerline{\includegraphics[scale=0.65]{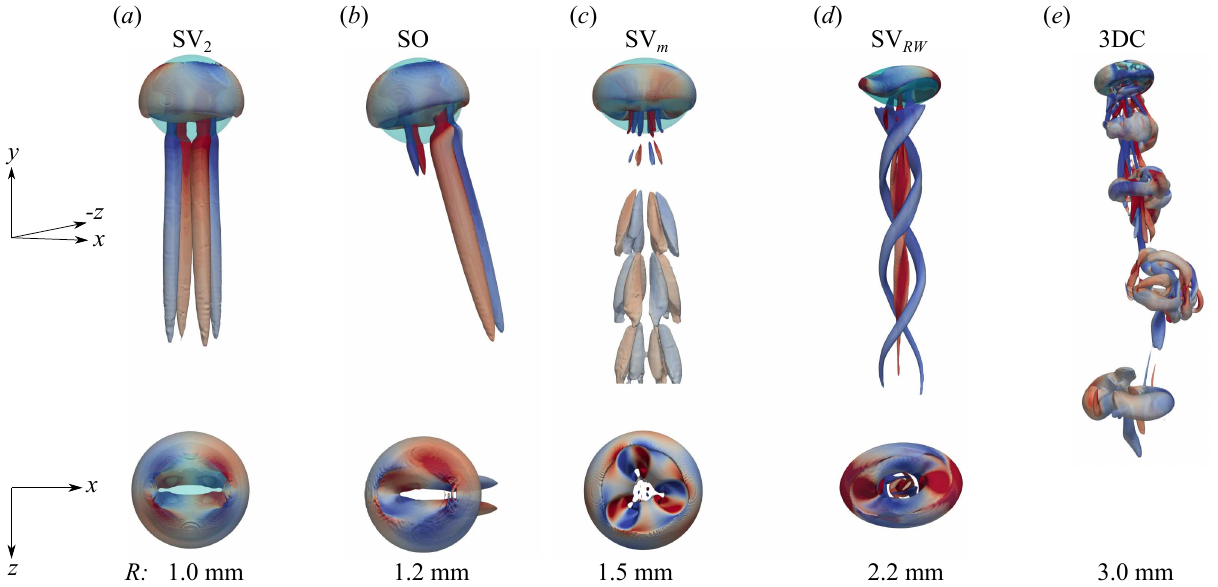}}
\caption{Wake structures associated with the five asymmetric regimes: $(a)$ $\text{SV}_2$ for $R=1.0\,\text{mm}$; $(b)$ SO for $R=1.2\,\text{mm}$; $(c)$ $\text{SV}_m$ for $R=1.5\,\text{mm}$, where the terminal disturbance is dominated by the $m=3$ mode; $(d)$ $\text{SV}_{RW}$ for $R=2.2\,\text{mm}$; $(e)$ 3DC for $R=3.0\,\text{mm}$. The wake is visualised using the $\lambda_2$ criterion, with iso-$\lambda_2$ surfaces coloured by the local value of $\overline{\omega}_y$; red and blue indicate negative and positive values, respectively, with the colour scale saturated at $|\overline{\omega}_y|=1.0$.}
\label{fig:l2_sum}
\end{figure}

In acceptable agreement with these findings, we observe that the axisymmetry of the flow can no longer be sustained once $R$ exceeds about $1.0\,\text{mm}$. Figure~\ref{fig:v_vs_R_3d} compares the terminal rise speed obtained in fully three-dimensional simulations with its axisymmetric counterpart, both computed in the present work. Starting from small $R$, the first regime encountered is the steady vertical regime ($\text{SV}_0$), associated with an axisymmetric flow field. As $R$ increases beyond about $1.0\,\text{mm}$, this base state is succeeded by a secondary steady vertical regime ($\text{SV}_2$), characterised by a lower rise speed in three dimensions. Figure~\ref{fig:l2_sum}$(a)$ visualizes the wake structure of a typical $\text{SV}_2$ state using the $\lambda_2$ criterion \citep{1995_Jeong}. The wake consists of four streamwise vortex threads of comparable intensity, indicating that the axisymmetry breaking is driven by an azimuthal mode with wavenumber $m=2$ \citep{ghidersa2000breaking,2025_Shi_drop}. Consequently, two orthogonal symmetry planes exist: one in which fluid elements are entrained towards the symmetry axis, and another in which they are driven away from it. In figure~\ref{fig:l2_sum}, these two planes are denoted by $x=0$ and $z=0$, respectively.

As $R$ increases beyond about $1.125\,\text{mm}$, the $\text{SV}_2$ regime is replaced by a steady oblique (SO) regime, similar to the first non-vertical paths observed for freely rising or settling solid spheres \citep{fabre2012steady,2010_Horowitz,2018_Auguste}. The reduction in rise speed is more pronounced than for the $\text{SV}_2$ regime. Figure~\ref{fig:l2_sum}$(b)$ shows the wake structure for a droplet with $R=1.2\,\text{mm}$, slightly above the $\text{SV}_2$--SO threshold. The wake still consists of four vortex threads, but the vortex pair located at lower $x$ is significantly weaker than the other pair, so that the flow field is now only uniplanar-symmetric. The latter vortex pair is known to generate a lift force towards $-x$ \citep{auton1987lift,legendre1998lift, adoua2009reversal}, thereby driving the lateral migration of the droplet in that direction.

As $R$ increases beyond about $1.5\,\text{mm}$, the SO regime is succeeded by a third steady vertical regime ($\text{SV}_m$), in which all droplets follow a $\text{SV}_0$ scenario except for $R=1.5\,\text{mm}$. For this specific case, the terminal rise speed is slightly smaller than its axisymmetric counterpart. Figure~\ref{fig:l2_sum}$(c)$ illustrates the corresponding terminal-state vortical structure. The $\lambda_2$ isosurfaces, coloured by the streamwise vorticity, reveal a clear threefold rotational symmetry about the rise axis, suggesting axisymmetry breaking through an azimuthal mode with wavenumber $m=3$ \citep{bonnefis2022stability}.

The $\text{SV}_m$ regime persists up to $R\approx2.05\,\text{mm}$, beyond which the system enters another asymmetric regime, hereafter referred to as the $\text{SV}_{RW}$ regime. In this regime, the droplet ultimately rises at a nearly time-independent speed, slightly smaller than its axisymmetric counterpart, while rotating steadily about the rise axis $y$. Figure~\ref{fig:l2_sum}$(d)$ illustrates the vortical structure associated with a typical $\text{SV}_{RW}$ state. The wake is characterised by a dominant axial vortex core accompanied by two co-rotating helical satellite threads. Remarkably, the two satellite threads have the same sign of $\overline{\omega}_y$, opposite to that of the core. This organization is consistent with a helical rotating-wave wake with even azimuthal symmetry, most plausibly $m=2$, and is reminiscent of the double-helical patterns reported in the wake of streamwise rotating spheres; see figure~11 of \citet{sierra2022unveiling}.

Beyond $R\approx2.3\,\text{mm}$, the $\text{SV}_{RW}$ regime is replaced by a highly asymmetric regime in which the flow enters a fully three-dimensional chaotic state, denoted as 3DC. In this regime, the steady self-spinning motion of the droplet can no longer be sustained. Figure~\ref{fig:l2_sum}$(e)$ shows the vortical structure for a typical 3DC case, $R=3.0\,\text{mm}$. No persistent wake symmetry can be identified; instead, vortices are shed repeatedly from the droplet surface. The corresponding rise speed oscillates at a frequency close to that found in the axisymmetric configuration, suggesting that shape oscillations play a central role in this regime.

Experimental data from \citet{wegener2009einfluss} are also included in figure~\ref{fig:v_vs_R_3d}. For small droplets, $R\leq1.0\,\text{mm}$, the measured rise speed is consistently lower than the present numerical prediction, with a relative deviation that becomes more pronounced as $R$ decreases. Since the external liquid is water, a polar liquid, this discrepancy may be attributed to contamination of the liquid--liquid system. Such contamination may partially immobilize the interface, thereby increasing the drag coefficient and reducing the rise speed \citep{2000_Magnaudet, kentheswaran2023impact}. This interpretation is supported by estimating the terminal rise speed from the drag correlation of \citet{2025_Shi_drop}, which applies to clean, i.e. surfactant-free, systems. Using the force balance in Eq.~\eqref{eq:drag_vs_buoy} for the case $R=0.5\,\text{mm}$ yields a dimensionless terminal rise speed $V_y=0.59$, in close agreement with the present numerical value $V_y=0.60$, but more than $10\%$ larger than the experimental value $V_y=0.50$. The possible influence of such contamination on the threshold of the $\text{SV}_0$--$\text{SV}_2$ transition is unknown, and might contribute to the approximately $15\%$ difference between the experimentally determined critical size, $R\approx1.15\,\text{mm}$, and the present prediction, $R\approx1.0\,\text{mm}$.

The influence of contamination appears to become much weaker for $R\gtrsim1.2\,\text{mm}$, where the measured rise speed differs from the present numerical result by less than $1\%$. As $R$ exceeds approximately $1.5\,\text{mm}$, the experimental data also show an abrupt increase in the rise speed. For $R>1.5\,\text{mm}$, they agree well with the present axisymmetric DNS up to $R\approx2.3\,\text{mm}$, beyond which the measured rise speed becomes smaller. Compared with the present fully three-dimensional results, the experimental values in the $\text{SV}_{RW}$ regime are consistently larger, especially at larger $R$. This mismatch is likely related to the shorter vertical distance over which the droplet was tracked in the experiment, approximately $400R$, compared with more than $900R$ in the present DNS. Indeed, as shown later in \S~\ref{sec:regime_sv_rw} , in a typical $\text{SV}_{RW}$ scenario the flow remains nearly axisymmetric until the vertical displacement exceeds approximately $500R$, after which the axial spinning motion develops and grows in time. 
In the 3DC regime, by contrast, the experimentally reported rise speed is significantly lower than that obtained in the present three-dimensional DNS. We expect this difference to be largely associated with the coexistence of another possible terminal state in this size range, as discussed in \S~\ref{sec:inf_dis}. This latter state may be more readily selected in experiments, owing to the practical release conditions of the droplet, which can introduce finite-amplitude disturbances and trigger transient shape oscillations.

Numerical results obtained by \citet{2015_Engberg} and \citet{charin2019dynamic} for $R$ up to $2.0\,\text{mm}$ are also shown in figure~\ref{fig:v_vs_R_3d}. The minimum grid size used by \citet{2015_Engberg} was approximately $\Delta_{\min}=R/48$, close to that employed in the present work, whereas that used by \citet{charin2019dynamic} was only about $\Delta_{\min}=R/13$, more than five times coarser. Likely owing to this difference, the agreement with the data of \citet{2015_Engberg} is within $3\%$, whereas deviations from the results of \citet{charin2019dynamic} are typically in the range $5$--$10\%$. Despite this moderate discrepancy, both previous studies confirm that, for $1.0\,\text{mm}\lesssim R<1.5\,\text{mm}$, the flow is asymmetric and characterised by either a biplanar- or a uniplanar-symmetric wake, corresponding to the $\text{SV}_2$ and SO regimes, respectively. The threshold droplet sizes marking the onset of these regimes were not specified in those studies. Both studies also reported an abrupt increase in the rise speed as $R$ approaches $1.5\,\text{mm}$, in line with the present three-dimensional DNS.

In the following subsections, we describe in more detail each of the asymmetric regimes encountered as $R$ increases.

\subsection{The $\text{SV}_2$ regime} \label{sec:regime_sv2}

\begin{figure}
\centerline{\includegraphics[scale=0.65]{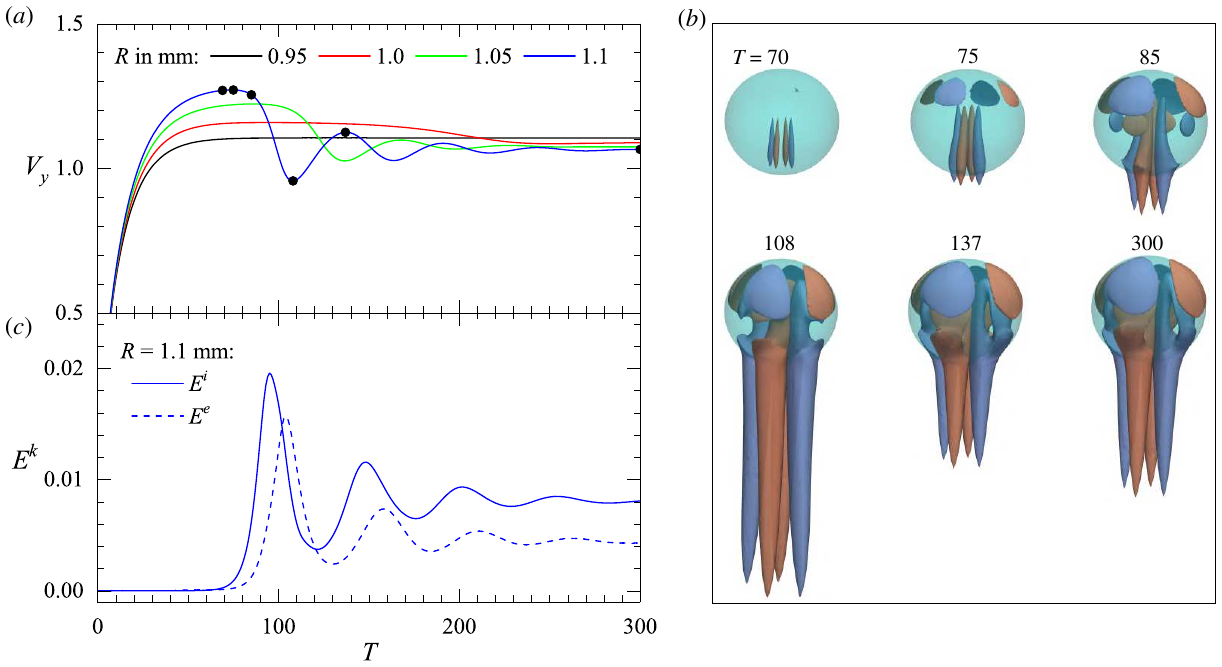}}
\caption{$(a)$ Rise speed as a function of time for $0.95\,\text{mm}\leq R \leq 1.1\,\text{mm}$. $(b)$ Isosurfaces of the axial vorticity, $\overline{\omega}_y = \pm 0.5$, at six selected time instants for $R=1.1\,\text{mm}$, indicated by numbers in $(b)$ and marked by circles in $(a)$. Blue and red threads correspond to positive and negative $\overline{\omega}_y$, respectively. $(c)$ Evolution of the internal and external azimuthal energies, $E^i$ and $E^e$, for $R=1.1\,\text{mm}$.}
\label{fig:v_vor_sv2}
\end{figure}

Figure~\ref{fig:v_vor_sv2}$(a)$ shows the time evolution of the droplet rise speed $V_y$ for $R$ increasing from $0.95$ to $1.1\,\text{mm}$. For all cases, $V_y$ first increases monotonically until it reaches a plateau. For $R=0.95\,\text{mm}$, this plateau is stable, and the corresponding three-dimensional flow field closely resembles that obtained in the axisymmetric configuration (not shown). For $R\geq1.0\,\text{mm}$, however, the plateau persists only over a finite time interval, whose duration decreases as $R$ increases. Beyond this interval, $V_y$ starts to decrease and exhibits several damped oscillations before reaching a terminal value lower than that of the initial plateau.

For $R\geq1.0\,\text{mm}$, the late-stage evolution of $V_y$ is closely linked to the development of the wake downstream of the droplet. Figure~\ref{fig:v_vor_sv2}$(b)$ shows isosurfaces of the axial vorticity, $\overline{\omega}_y$, at selected time instants for $R=1.1\,\text{mm}$. Before $V_y$ starts to decrease, namely at $T=70$, two pairs of counter-rotating streamwise vortices have already formed inside the droplet, indicating that the axisymmetry of the internal flow has already broken down. These vortex threads grow in time and, at $T=75$, generate, through the interfacial boundary conditions, four external counterparts of the same signs. This is also the stage at which $V_y$ starts to decrease. Once formed, the external vortex threads grow rapidly, reaching a maximum downstream extension at $T=108$, after which they contract and subsequently re-extend. This extension--contraction cycle is repeated several times before saturation. Since the vortex cores are low-pressure regions, their presence increases the pressure difference between the front and rear of the droplet, compared with the axisymmetric state, and hence enhances the drag at a given $V_y$. As drag and buoyancy remain nearly balanced, $V_y$ reaches a local minimum shortly after the vortex threads attain their maximum downstream extension, for instance at $T=108$, whereas the opposite occurs when the vortex threads contract to a local minimum, for instance at $T=137$.

The fact that, during the early stages of axisymmetry breaking, the axial vorticity, and hence the disturbance, is significant only inside the droplet suggests that the $\text{SV}_0$--$\text{SV}_2$ transition is driven by an internal flow instability \citep{2024_Shi_drop,2025_Shi_drop,2026_gode}. To confirm this, we monitor the evolution of the perturbation energy. A convenient measure is the mean kinetic energy of the azimuthal velocity component, hereafter referred to as the azimuthal energy \citep{thompson2001kinematics,2007_Magnaudet,2025_Shi_drop}, defined as
\begin{equation}
E^k=\frac{1}{\rho^e \mathcal{V} \left(gR\right)} \int_{\mathcal{V}^k}{\rho^k u_\theta^2\,\mathrm{d}\mathcal{V}^k},
\label{eq:azi_eng}
\end{equation}
where $E^k$, with $k=i$ or $e$, denotes the normalised azimuthal energy inside or outside the droplet, respectively. Figure~\ref{fig:v_vor_sv2}$(c)$ shows the evolution of the internal and external azimuthal energies for $R=1.1\,\text{mm}$. The axisymmetry of the base flow breaks down at $T\approx65$, as indicated by the onset of growth of the internal azimuthal energy $E^i$, whereas the external azimuthal energy $E^e$ remains negligible until $T\approx75$. Afterwards, both energy components undergo underdamped oscillations, with the phase of $E^e$ consistently lagging behind that of $E^i$.

\begin{figure}
\centerline{\includegraphics[scale=0.6]{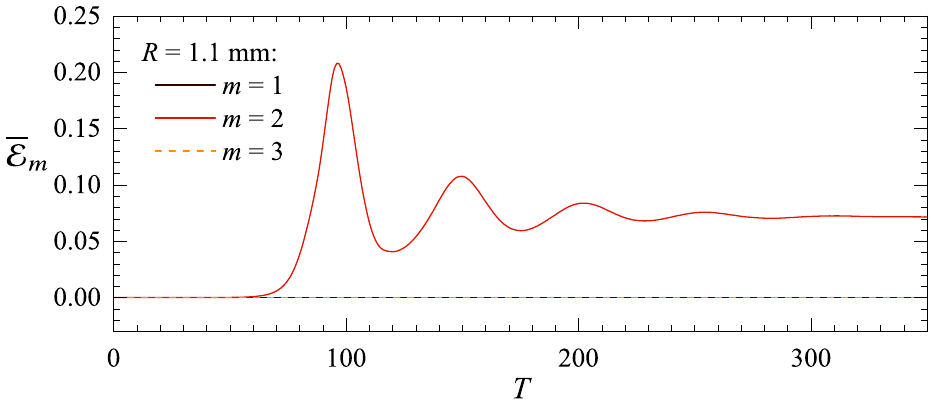}}
\caption{Evolution of the normalised modal energy associated with azimuthal wavenumber $m$, $\overline{\mathcal{E}}_m\equiv \mathcal{E}_m/(gR)$, during a $\text{SV}_2$ scenario for $R=1.1\,\text{mm}$. The definition of $\mathcal{E}_m$ is given in Eq.~\eqref{eq:app_Em_cont}.}
\label{fig:mode_sv2}
\end{figure}

The biplanar-symmetric vortical structure shown in figure~\ref{fig:v_vor_sv2}$(b)$ suggests that the $\text{SV}_0$--$\text{SV}_2$ transition is associated with an azimuthal mode of wavenumber $m=2$. To confirm this, we perform a two-dimensional azimuthal Fourier analysis of the flow inside the droplet in the cross-stream plane passing through the droplet centroid. The technical details are provided in Appendix~\ref{app:fft}; only the main definitions are recalled here. Let $q(r,\theta,t)$ denote a scalar field to be analysed in the cross-stream plane, where $(r,\theta)$ are the radial and azimuthal coordinates. For each fixed radius $r$, the azimuthal Fourier coefficient of wavenumber $m$ is defined as
\begin{equation}
\hat{q}_m(r,t)=\frac{1}{2\pi}\int_0^{2\pi} q(r,\theta,t)\,\exp(-\mathrm{i}m\theta)\,\mathrm{d}\theta\,.
\label{eq:app_qhat_cont}
\end{equation}
The modal energy associated with azimuthal wavenumber $m$ is then defined as
\begin{equation}
\mathcal{E}_m(t)=\int_{r_{\min}}^{r_{\max}} |\hat{q}_m(r,t)|^2\,r\,\mathrm{d}r\,,
\label{eq:app_Em_cont}
\end{equation}
where $r_{\min}$ and $r_{\max}$ denote the lower and upper bounds of the radial interval used in the analysis. Since our main interest lies in the internal flow field, we set $r_{\min}=0.05R$ and $r_{\max}=0.95R$; the finite lower bound avoids the coordinate singularity at $r=0$. The scalar field $q$ is taken as the streamwise vorticity, defined as $\omega_v=\boldsymbol{\omega}\cdot\boldsymbol{e}_v$, where $\boldsymbol{e}_v\equiv \boldsymbol{v}/|\boldsymbol{v}|$ is the unit vector aligned with the instantaneous droplet velocity. With this definition, $\omega_v$ reduces to the axial vorticity $\omega_y$ when the path is vertical. The normalised modal energy, $\overline{\mathcal{E}}_m\equiv \mathcal{E}_m/(gR)$, is used to identify the dominant azimuthal wavenumber and monitor its temporal evolution.

Figure~\ref{fig:mode_sv2} shows the evolution of $\overline{\mathcal{E}}_m$ for $m\leqslant3$. As expected, the initial growth is dominated by the $m=2$ mode, with $\overline{\mathcal{E}}_2$ ultimately saturating at a time-independent value. By contrast, the modal energies associated with $m=1$ and $m=3$ remain negligible throughout the evolution. This confirms the dominant role of the $m=2$ mode in the $\text{SV}_0$--$\text{SV}_2$ transition.

\subsection{The SO regime} \label{sec:regime_so}

\begin{figure}
\centerline{\includegraphics[scale=0.65]{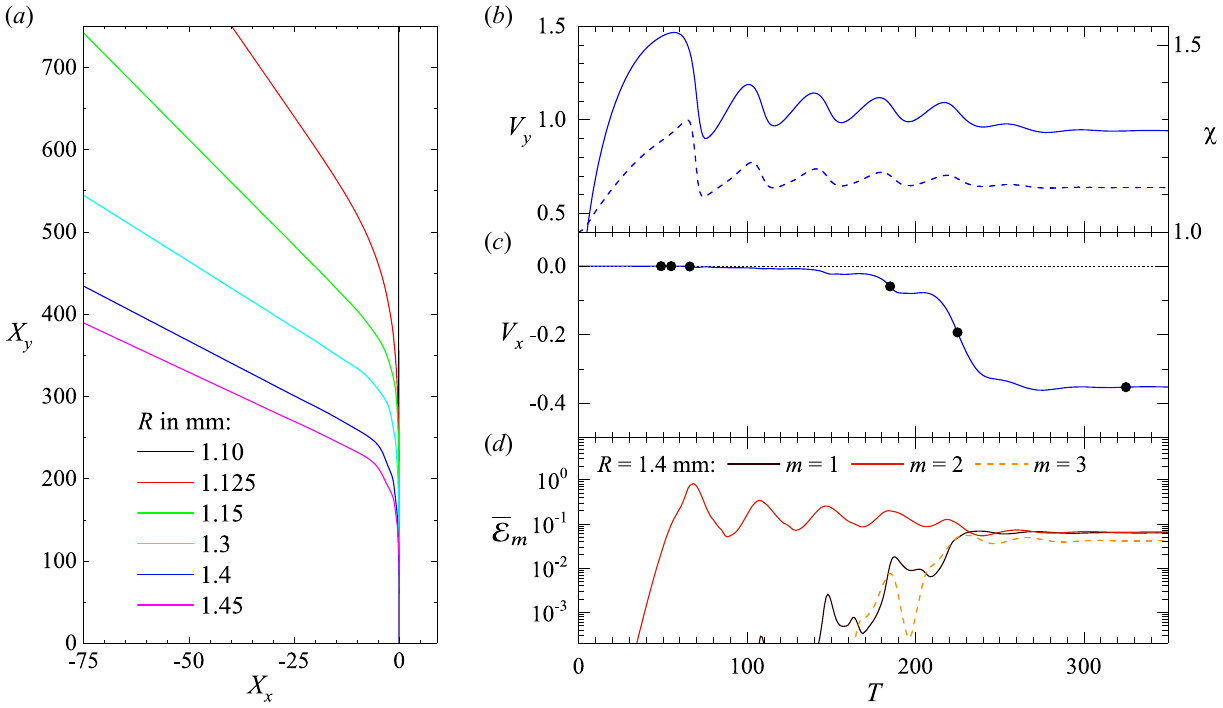}}
\caption{$(a)$ Rise paths of droplets for $1.10\,\text{mm}\leq R \leq 1.45\,\text{mm}$. $(b)$--$(d)$ Evolution of several characteristics of the droplet motion during an SO scenario for $R=1.4\,\text{mm}$. $(b)$ Rise speed $V_y$ (solid line, left axis) and aspect ratio $\upchi$ (dashed line, right axis); $(c)$ horizontal velocity $V_x$; $(d)$ normalised modal energies.}
\label{fig:v_rise_so}
\end{figure}

Figure~\ref{fig:v_rise_so}$(a)$ shows the droplet rise paths for $1.10\,\text{mm}\leq R \leq 1.45\,\text{mm}$. As $R$ increases from $1.1$ to $1.125\,\text{mm}$, the terminal path transitions from vertical to steady oblique, similar to what is observed for spherical particles rising or settling freely under buoyancy or gravity \citep{2004_Jenny,veldhuis2009freely,2010_Horowitz,2018_Auguste}. In the latter case, however, the $m=2$ mode remains stable and only the $m=1$ mode becomes active during the axisymmetry-breaking process \citep{fabre2012steady}. Consequently, the first non-vertical path there corresponds to an $\text{SV}_0$--SO transition, whereas the present case involves an $\text{SV}_2$--SO transition. Figure~\ref{fig:v_rise_so}$(a)$ also shows that the vertical position at which the lateral migration sets in decreases slightly as $R$ increases. Once the lateral motion has saturated, the drift angle of the rise path with respect to the vertical increases from $8.3^\circ$ to $22.7^\circ$ as $R$ increases from $1.125$ to $1.45\,\text{mm}$.

\begin{figure}
\centerline{\includegraphics[scale=0.65]{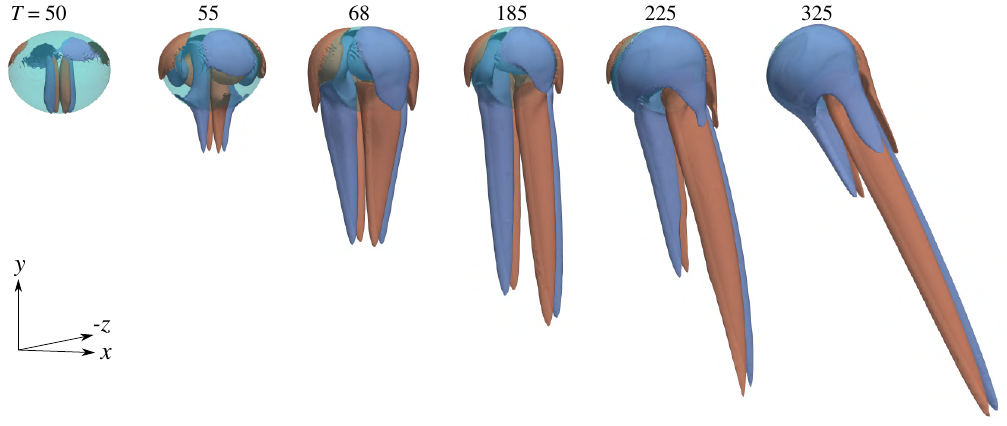}}
\caption{Isosurfaces of the axial vorticity, $\overline{\omega}_y=\pm0.5$, at six selected time instants, indicated by numbers at the top of each panel and marked by circles in figure~\ref{fig:v_rise_so}$(c)$. Blue and red threads correspond to positive and negative $\overline{\omega}_y$, respectively.}
\label{fig:vor_so}
\end{figure}

Figures~\ref{fig:v_rise_so}$(b)$--$(d)$ show the evolution of several characteristics of the droplet motion for $R=1.4\,\text{mm}$. The corresponding $\overline{\omega}_y$ isosurfaces at selected instants are displayed in figure~\ref{fig:vor_so}. Before the lateral motion sets in, say for $T\lesssim140$, the evolution closely resembles that of a $\text{SV}_2$ scenario, such as that observed for $R=1.1\,\text{mm}$. In particular, the rise speed, as well as the internal and external azimuthal energies (not shown), undergoes damped oscillations. The $\overline{\omega}_y$ isosurfaces confirm that the asymmetry is initially generated and amplified inside the droplet (see the inset at $T=50$ in figure~\ref{fig:vor_so}), and that the flow structure following the first bifurcation is biplanar-symmetric.

In figure~\ref{fig:v_rise_so}$(c)$, the magnitude of the horizontal velocity becomes larger than $0.01$ at $T\approx140$, indicating that the flow no longer preserves the biplanar symmetry established after the first bifurcation. Indeed, as the $\overline{\omega}_y$ isosurfaces for $T\geq185$ show, the wake then develops in a left--right asymmetric manner. As time proceeds, the left vortex pair shrinks, while the right pair extends further downstream, so that in the terminal state the intensity of the right pair is significantly larger than that of the left pair. The saturated wake structure differs from that associated with a spherical particle following a steady oblique path; see, for instance, figure~17 of \citet{pan2019wake}. Although both cases correspond to a steady oblique path, the wake in the latter case consists of a single pair of streamwise vortices, since only the $m=1$ mode remains active \citep{fabre2012steady}. By contrast, in the present SO regime, the coexistence of two vortex pairs of unequal intensity suggests that both the $m=1$ and $m=2$ modes remain active in the saturated state.

This interpretation is supported by figure~\ref{fig:v_rise_so}$(d)$, which shows the evolution of the normalised modal energies. As expected, the primary bifurcation associated with the $\text{SV}_0$--$\text{SV}_2$ transition is triggered by the $m=2$ mode. Beyond $T\approx140$, i.e. once $|V_x|$ first becomes non-negligible, the $m=1$ mode starts to grow, leading to the $\text{SV}_2$--SO transition. The relative intensity of the $m=1$ mode in the saturated state increases with $R$: the ratio $\overline{\mathcal{E}}_1/\overline{\mathcal{E}}_2$ increases from approximately $12.5\%$ to nearly unity as $R$ increases from $1.125$ to $1.45\,\text{mm}$. Finally, nonlinear interactions between the first two modes may generate a higher-order mode with $m=3$, whose presence is also detected in the Fourier decomposition.

Before moving to the next subsection, we briefly discuss the time-dependent deformation of the droplet during the transition. Figure~\ref{fig:v_rise_so}$(b)$ shows, using the dashed line and the right axis, the evolution of the aspect ratio $\upchi$ for the droplet with $R=1.4\,\text{mm}$. Clearly, the variations of $\upchi$ closely follow those of the rise speed $V_y$: $\upchi$ reaches a local maximum shortly after $V_y$ reaches a peak, and vice versa. This behaviour differs from that observed in the axisymmetric configuration (see figure~\ref{fig:v_chi_large_R}$c$), where a maximum of $\upchi$ corresponds to a local minimum of $V_y$. In the present SO scenario, the shape variation is a response to changes in the dynamic pressure around the droplet, and hence to changes in $V_x^2+V_y^2$, rather than the signature of an intrinsic shape instability as in the axisymmetric oscillatory regime.

\subsection{The $\text{SV}_m$ regime} \label{sec:regime_svm}

\begin{figure}
\centerline{\includegraphics[scale=0.65]{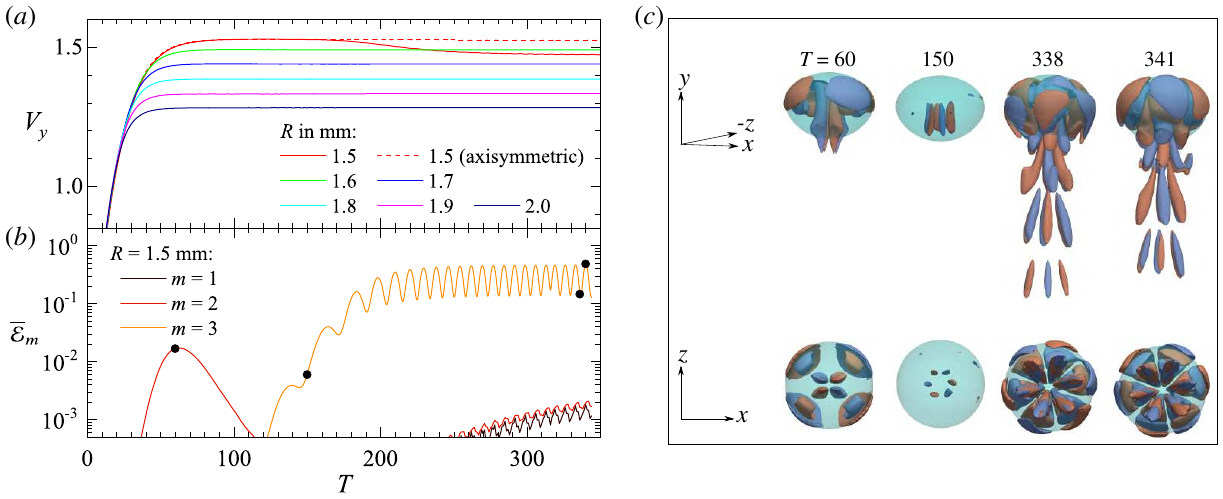}}
\caption{$(a)$ Rise speed as a function of time for $1.5\,\text{mm}\leq R \leq 2.0\,\text{mm}$. $(b)$ Evolution of the normalised modal energy associated with azimuthal wavenumber $m$ during an $\text{SV}_m$ scenario for $R=1.5\,\text{mm}$. $(c)$ Isosurfaces of the axial vorticity, $\overline{\omega}_y=\pm0.25$, at selected time instants, indicated by numbers at the top of each panel and marked by circles in $(b)$. Blue and red threads correspond to positive and negative $\overline{\omega}_y$, respectively.}
\label{fig:v_mode_svm}
\end{figure}

As $R$ exceeds about $1.5\,\text{mm}$, the terminal path switches from steady oblique back to steady vertical. Figure~\ref{fig:v_mode_svm}$(a)$ shows the evolution of the droplet rise speed for $1.5\,\text{mm}\leq R \leq 2.0\,\text{mm}$. The corresponding horizontal velocity components are not shown, as they remain vanishingly small throughout the evolution. For all cases (except one), $V_y$ increases monotonically before reaching a terminal value equal to that obtained in the axisymmetric configuration. The only exception is the case $R=1.5\,\text{mm}$. Compared with its axisymmetric counterpart, shown by the red dashed line, the three-dimensional case undergoes a moderate reduction of $V_y$ during the late stage of the evolution and eventually stabilizes at a terminal value approximately $4\%$ smaller than the axisymmetric one.

Azimuthal Fourier analysis, combined with visualisation of the vortical structure, provides insight into the flow evolution for $R=1.5\,\text{mm}$. As shown in figure~\ref{fig:v_mode_svm}$(b)$, before $V_y$ reaches its first plateau, the familiar $m=2$ mode starts to grow, as also indicated by the vortical structure at $T=60$ in figure~\ref{fig:v_mode_svm}$(c)$. However, this mode grows only briefly and then decays towards a negligible level. Beyond $T\approx130$, the axisymmetry of the flow breaks down again, now through an azimuthal mode with wavenumber $m=3$. The vortical structure at $T=150$ suggests that the initial growth of this mode takes place within the droplet, and that this mode may therefore also be associated with an internal flow instability, like the $m=2$ mode discussed above.

This $m=3$ mode differs from that observed for $R=1.4\,\text{mm}$ (see figure~\ref{fig:v_rise_so}$d$). In the present case, it arises as a primary instability of the system, whereas for smaller droplets it results from nonlinear interactions between the $m=1$ and $m=2$ modes. Returning to figure~\ref{fig:v_mode_svm}$(b)$, the $m=3$ mode continues to grow until the system reaches a fully developed time-dependent state, characterised by regular oscillations of $\overline{\mathcal{E}}_3(t)$, and also of $V_y(t)$, although with a vanishingly small amplitude. The corresponding dimensionless frequency is $\overline{f}\approx0.17$, less than half the frequency associated with the axisymmetric \q{mode-2} shape oscillations of a nearly spherical droplet \citep{1932_Lamb,lalanne2013effect}. The last two snapshots in figure~\ref{fig:v_mode_svm}$(c)$ show that the signs of the three vortex pairs associated with the $m=3$ mode reverse every half-period. Finally, we note that the brief growth of the $m=2$ mode is a general feature of cases with $1.5\,\text{mm}\leq R\leq2.0\,\text{mm}$, although its peak modal energy decreases as $R$ increases. By contrast, the onset of a time-dependent $m=3$ mode is observed only for $R=1.5\,\text{mm}$.

\subsection{The $\text{SV}_{RW}$ regime} \label{sec:regime_sv_rw}

\begin{figure}
\centerline{\includegraphics[scale=0.675]{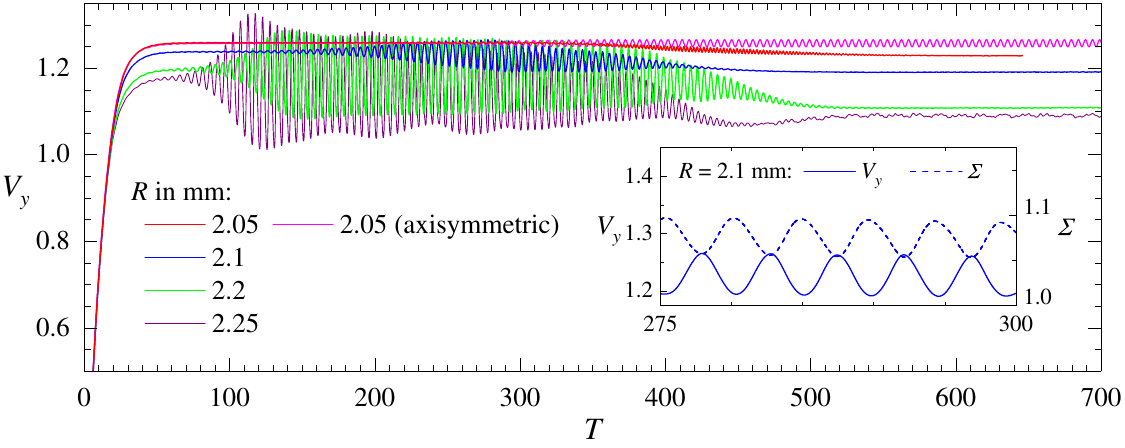}}
\caption{Evolution of the rise speed of droplets with $2.05\,\text{mm}\leq R\leq2.25\,\text{mm}$. The inset shows, over a selected time interval, the simultaneous evolution of the rise speed $V_y$ and the normalised interfacial area $\it\Sigma$ for the droplet with $R=2.1\,\text{mm}$.}
\label{fig:v_rise_sv_rw}
\end{figure}

The flow axisymmetry breaks down again as $R$ increases beyond a critical value between $2.0$ and $2.05\,\text{mm}$. Figure~\ref{fig:v_rise_sv_rw} shows the evolution of the rise speed for droplets with $2.05\,\text{mm}\leq R\leq2.25\,\text{mm}$. At $R=2.05\,\text{mm}$, the rise speed in the axisymmetric configuration eventually undergoes regular oscillations. For the same droplet in the fully three-dimensional configuration, these oscillations are visible only over an intermediate time interval and start to decay once $T$ exceeds approximately $300$. After this decay, $V_y$ ultimately stabilizes at a time-independent value slightly smaller than its mean axisymmetric counterpart. This behaviour persists as $R$ increases, except that the relative reduction in the terminal value of $V_y$ becomes slightly larger, increasing from approximately $3\%$ at $R=2.05\,\text{mm}$ to approximately $8\%$ at $R=2.25\,\text{mm}$. The inset of figure~\ref{fig:v_rise_sv_rw} shows, over a selected time interval, the simultaneous evolution of the dimensionless interfacial area $\it\Sigma$ and the rise speed for the three-dimensional case at $R=2.1\,\text{mm}$. Clearly, $V_y$ reaches a local minimum when $\it\Sigma$, and hence the deformation level of the droplet, reaches a maximum, and vice versa. This indicates that the rise-speed oscillations observed during the intermediate stage are primarily caused by droplet-shape oscillations.

\begin{figure}
\centerline{\includegraphics[scale=0.65]{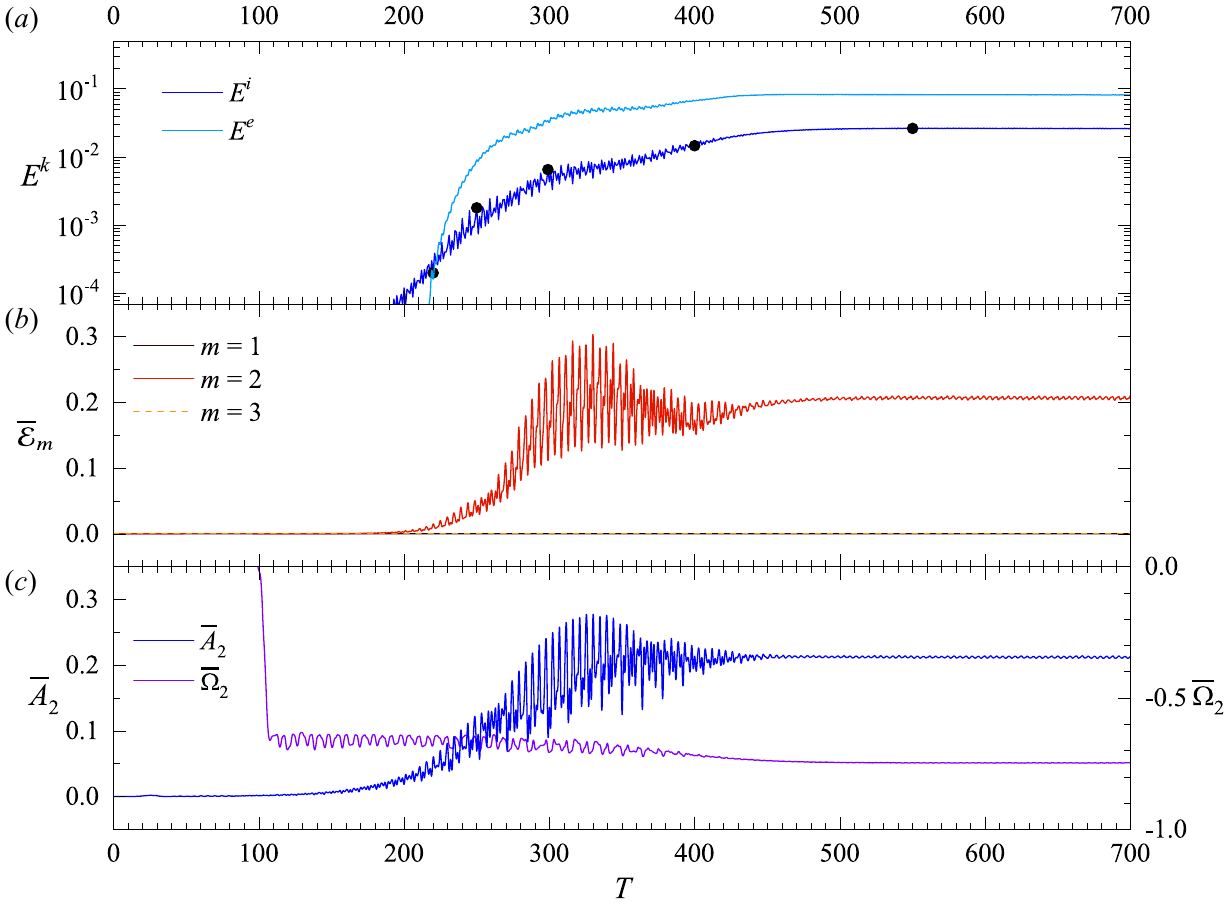}}
\caption{Evolution of quantities characterising the axisymmetry breaking for the droplet with $R=2.1\,\text{mm}$. $(a)$ Internal ($E^i$) and external ($E^e$) azimuthal energies; $(b)$ normalised modal energies, defined in Eqs.~(\ref{eq:app_qhat_cont}--\ref{eq:app_Em_cont}); $(c)$ global modal amplitude, left axis, and angular velocity, right axis, of the disturbance associated with azimuthal wavenumber $m=2$.}
\label{fig:mode_sv_rw}
\end{figure}

\begin{figure}
\centerline{\includegraphics[scale=0.65]{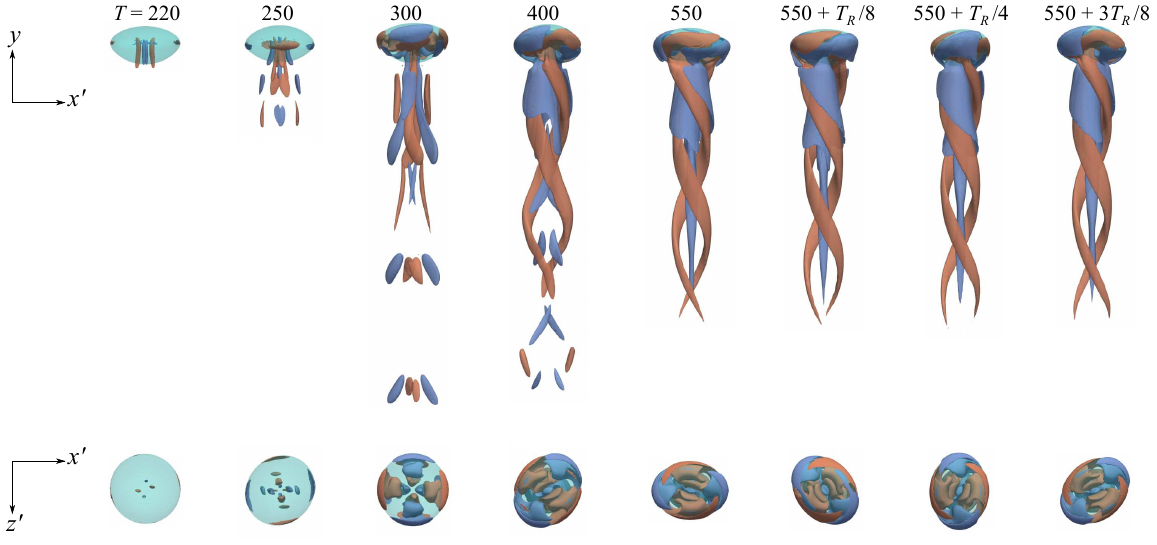}}
\caption{Isosurfaces of the axial vorticity, $\overline{\omega}_y=\pm0.75$, for the droplet with $R=2.1\,\text{mm}$ at selected time instants, indicated by numbers at the top of each panel and, for the first five instants, marked by circles in figure~\ref{fig:mode_sv_rw}$(a)$. First row: front view; second row: top view. The plane $(x^\prime,z^\prime)$ is horizontal, with the $x^\prime$ and $z^\prime$ axes selected arbitrarily.}
\label{fig:vor_sv_rw}
\end{figure}

Let us focus on the case $R=2.1\,\text{mm}$ and examine the evolution of several quantities characterising the axisymmetry breaking of the system. Figure~\ref{fig:mode_sv_rw}$(a)$ shows the evolution of the internal ($E^i$) and external ($E^e$) azimuthal energies. The first detectable departure from axisymmetry occurs around $T\approx200$ within the droplet, as indicated by the initial growth of $E^i$. The external azimuthal energy $E^e$ starts to increase shortly afterwards and rapidly exceeds $E^i$, showing that the disturbance is then transmitted to the surrounding liquid and amplified in the wake. Figure~\ref{fig:mode_sv_rw}$(b)$ shows the corresponding evolution of the normalised modal energies. As in the $\text{SV}_0$--$\text{SV}_2$ transition, the axisymmetry breaking is associated with an azimuthal mode of wavenumber $m=2$. Here, however, the corresponding modal energy first undergoes large-amplitude oscillations before saturating, a behaviour absent during the $\text{SV}_0$--$\text{SV}_2$ transition. These oscillations are likely linked to the underlying shape oscillations discussed above. They gradually vanish in the fully developed state, in which $\overline{\mathcal{E}}_2$ reaches a nearly time-independent value.

Figure~\ref{fig:vor_sv_rw} shows the $\overline{\omega}_y$ isosurfaces at several selected instants, marked by circles in figure~\ref{fig:mode_sv_rw}$(a)$ for the first five snapshots. During the initial stage of axisymmetry breaking, at $T=220$, two pairs of streamwise vortices are generated and amplified inside the droplet, similar to what is observed during the $\text{SV}_0$--$\text{SV}_2$ transition. As the disturbance propagates outwards, however, the development of a biplanar-symmetric wake is less efficient than in the latter case. More specifically, during each cycle of the $V_y$ oscillations, and hence of the shape oscillations, the vorticity generated near the droplet interface detaches from the droplet, leaving the near-wake region almost irrotational; see the snapshots at $T=250$ and $T=300$. This shedding process starts to weaken when $T$ exceeds approximately $350$, at which point the amplitude of the $V_y$ oscillations begins to decay, allowing the wake to develop over a sizeable distance downstream (see the snapshot at $T=400$).

In the fully developed state, illustrated by the snapshot at $T=550$, the shedding process has disappeared. The droplet adopts a frozen, egg-like shape and rises steadily upwards. A closer inspection of the vortical structure indicates that the flow undergoes an approximately steady azimuthal drift with angular velocity $2\pi/T_R$, where $T_R\approx8.4$ is the period of this azimuthal rotation. The last four snapshots of figure~\ref{fig:vor_sv_rw} show the evolution of the vortical structure over a time interval of $3T_R/8$ in a frame translating with the droplet. They make clear that the wake is nearly stationary in a reference frame rotating about the vertical axis through the droplet centroid. The wake consists of three main vortex threads: a dominant thread concentrated near the rise axis and extending almost straight downstream, and two secondary threads. The latter have the same sign, opposite to that of the dominant thread, and wrap around it successively. 

The spiral organisation of the terminal wake structure shown in figure~\ref{fig:vor_sv_rw} is similar to that observed behind spiralling bubbles \citep{mougin2002path,2016_Cano-Lozano}, although the dominant azimuthal wavenumber differs, being $m=1$ in the latter case and $m=2$ in the present one. For spiralling bubbles, the wake and the associated path correspond to a rotating-wave (RW) solution emerging from the Hopf bifurcation responsible for the first path instability \citep{fabre2008bifurcations,tchoufag2015weakly,2024_Bonnefis}. To assess whether the dominant mode identified here has the same character, we examine its global modal amplitude and angular velocity. For this purpose, we introduce a global complex Fourier coefficient by integrating radially the local azimuthal Fourier coefficient defined in \S~\ref{sec:regime_sv2}, namely
\begin{equation}
Q_m(t)=\int_{r_{\min}}^{r_{\max}} \hat{q}_m(r,t)\,r\,\mathrm{d}r,
\label{eq:global_Qm}
\end{equation}
where $\hat{q}_m(r,t)$ is the azimuthal Fourier coefficient of wavenumber $m$ defined in Eq.~\eqref{eq:app_qhat_cont}. Its modulus,
\begin{equation}
A_m(t)=|Q_m(t)|,
\label{eq:global_Am}
\end{equation}
defines the global modal amplitude, while its phase,
\begin{equation}
\Phi_m(t)=\arg\left(Q_m(t)\right),
\label{eq:global_phase}
\end{equation}
yields the angular velocity
\begin{equation}
\Omega_m(t)=\frac{1}{m}\frac{\mathrm{d}\Phi_m}{\mathrm{d}t}.
\label{eq:global_omega}
\end{equation}
For a $RW$ state, the flow is steady in a frame rotating with the wave and therefore appears in the laboratory frame as a constant-amplitude structure with a uniformly drifting phase. The corresponding dominant global Fourier coefficient may then be written as
\begin{equation}
Q_m(t)\approx A_m \exp(-\mathrm{i}m\Omega t),
\label{eq:rw_qm}
\end{equation}
with $A_m$ approximately constant and $\Phi_m(t)$ varying linearly in time \citep{tchoufag2015weakly}. Figure~\ref{fig:mode_sv_rw}$(c)$ shows the evolution of the normalised modal amplitude $\overline{A}_2=A_2/(g^{1/2}R^{3/2})$ and angular velocity $\overline{\Omega}_2=\Omega_2/(g^{1/2}R^{-1/2})$. Although both quantities exhibit some scatter during the transient stage, they eventually approach nearly time-independent values. This behaviour is consistent with a $RW$ state associated with azimuthal wavenumber $m=2$. The terminal angular velocity yields a rotation period $T_R=2\pi/|\Omega_2|=8.4$, in agreement with that inferred directly from the evolution of the vortical structure. The same behaviour is observed for all droplets with $2.05\,\text{mm}\leq R\leq2.25\,\text{mm}$, for which we find $T_R\approx8.4\pm0.1$, with only a weak dependence on $R$.

\subsection{The 3DC regime} \label{sec:regime_3dc}

As $R$ exceeds approximately $2.3\,\text{mm}$, the $\text{SV}_{RW}$ regime is succeeded by a fully three-dimensional chaotic regime, referred to as 3DC.

\begin{figure}
\centerline{\includegraphics[scale=0.65]{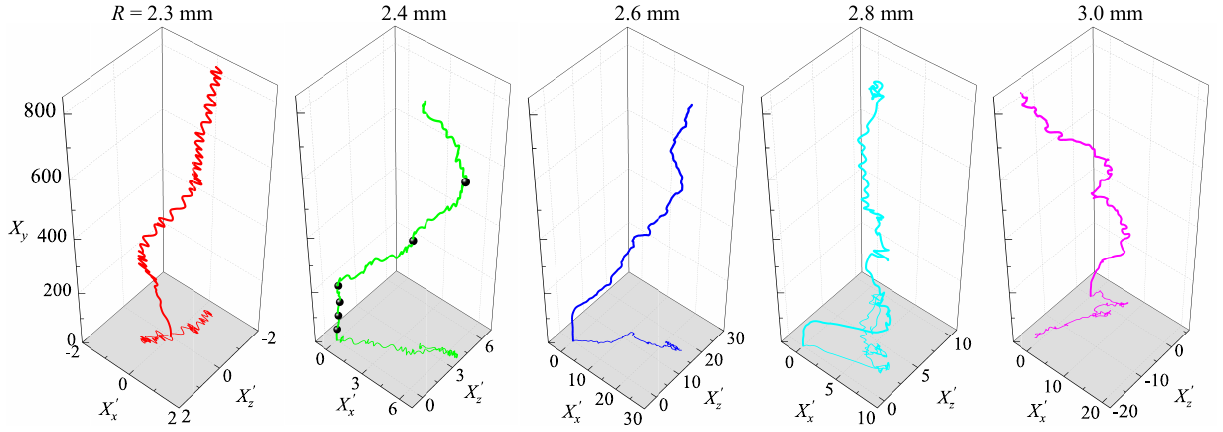}}
\caption{Three-dimensional paths in the 3DC regime for $2.3\,\text{mm}\leq R \leq 3.0\,\text{mm}$.}
\label{fig:3DC_path}
\end{figure}

Figure~\ref{fig:3DC_path} shows the trajectories in the 3DC regime for $R$ increasing from $2.3$ to $3.0\,\text{mm}$. For $R\leq2.6\,\text{mm}$, the trajectory is only weakly three-dimensional: its projection onto the horizontal plane extends preferentially along a single lateral direction. More specifically, within this size range, the droplet first rises vertically until its vertical displacement exceeds a threshold, typically $100$--$200R$, that decreases with increasing $R$. It then starts to migrate laterally along a preferred direction. At later stages, the direction of lateral migration may reverse temporarily, but the path remains largely confined to the same vertical plane. In this weakly 3DC regime, the total lateral excursion over a vertical displacement of approximately $900R$ increases with $R$, growing from about $2.25R$ at $R=2.3\,\text{mm}$ to about $30R$ at $R=2.6\,\text{mm}$. The situation changes for $R\gtrsim2.8\,\text{mm}$. For these larger droplets, the initial evolution is similar: the droplet first rises vertically and then migrates laterally along a preferred direction. However, once the lateral displacement exceeds a distance of $\mathcal{O}(R)$, which decreases with increasing $R$, the path becomes fully three-dimensional. It then exhibits abrupt changes of direction, and its horizontal projection tends to densely fill a finite region of the cross-stream plane. The amplitude of the horizontal displacement during this fully three-dimensional stage is about $10R$ at $R=2.8\,\text{mm}$ and increases to about $20R$ at $R=3.0\,\text{mm}$.

\begin{figure}
\centerline{\includegraphics[scale=0.65]{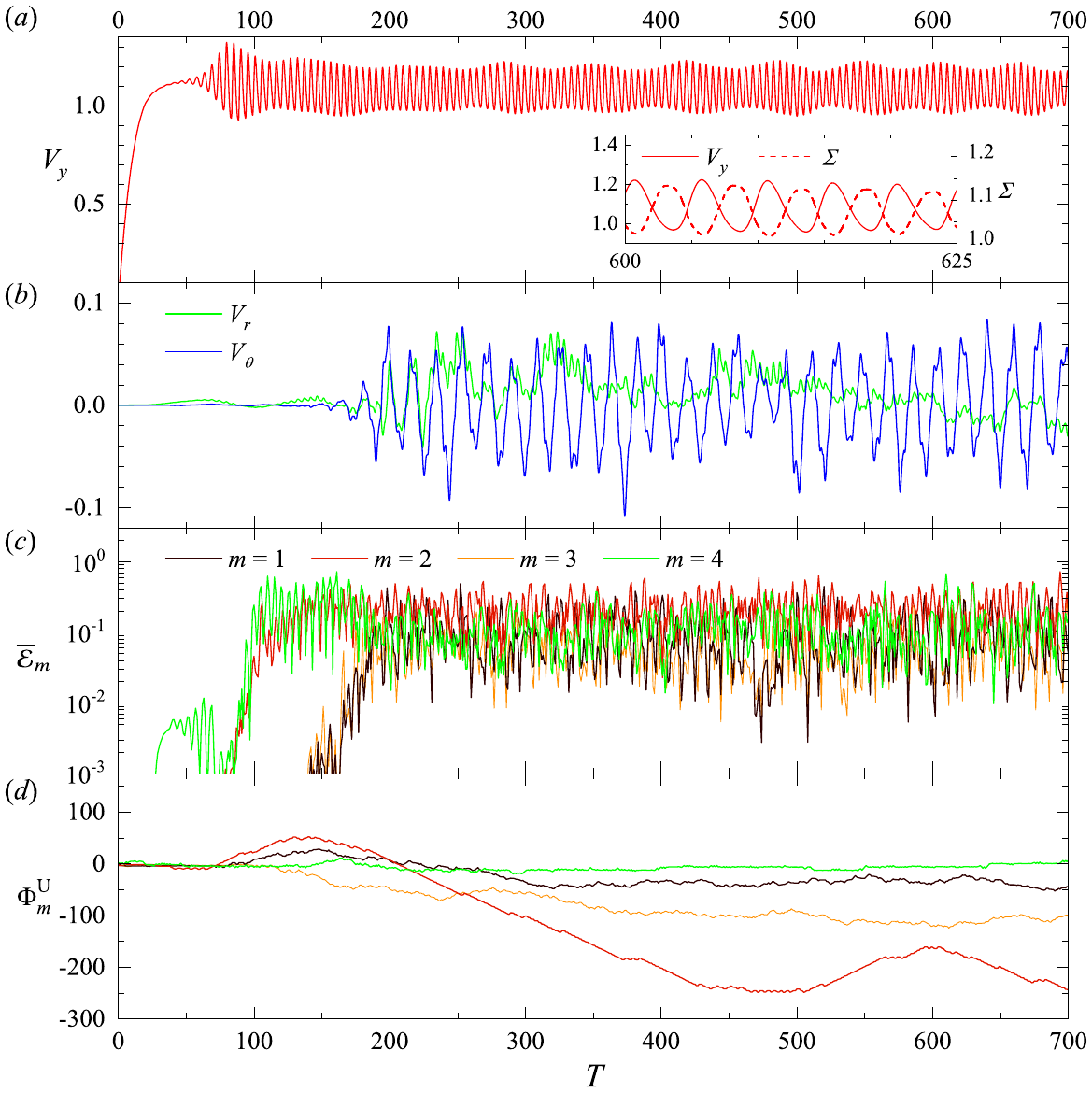}}
\caption{Evolution of quantities characterising droplet motion and flow structure during a 3DC scenario for $R=2.4\,\text{mm}$. $(a)$ Rise speed $V_y$; the inset shows the simultaneous evolution of $V_y$ and the dimensionless interfacial area $\it\Sigma$ over a selected time interval. $(b)$ Radial and azimuthal components of the horizontal velocity, $V_r$ and $V_\theta$. $(c)$ Normalised modal energies associated with the various azimuthal wavenumbers. $(d)$ Unwrapped phase angle $\Phi_m^{\text{U}}$ of the corresponding global Fourier coefficients.}
\label{fig:motion_r2.4}
\end{figure}

To gain further insight into the motion in the 3DC regime, figure~\ref{fig:motion_r2.4} illustrates the evolution of several quantities for the case $R=2.4\,\text{mm}$. Figure~\ref{fig:motion_r2.4}$(a)$ shows the time history of the rise speed. Unlike in a $\text{SV}_{RW}$ scenario, where the rise speed ultimately stabilizes at a time-independent value, here $V_y$ continues to oscillate even at long time. The oscillation amplitude is about $0.12\pm0.02$, approximately $65\%$ of that obtained in the corresponding axisymmetric configuration. The mean rise speed and the oscillation frequency are $V_m=1.08$ and $\overline{f}=0.19$, respectively, both close to their axisymmetric counterparts. The inset of figure~\ref{fig:motion_r2.4}$(a)$ shows the simultaneous evolution of the interfacial area $\it\Sigma$ and the rise speed over a selected time interval. As in the transient stage of a $\text{SV}_{RW}$ scenario (see the inset of figure~\ref{fig:v_rise_sv_rw}), $V_y$ reaches a local minimum when $\it\Sigma$ reaches a maximum, and vice versa, indicating that the rise-speed oscillations are associated with droplet-shape oscillations.

Figure~\ref{fig:motion_r2.4}$(b)$ shows the evolution of the radial and azimuthal velocity components, $V_r$ and $V_\theta$, in the horizontal plane. These two components remain vanishingly small until $T\approx150$, beyond which they reach finite values and start to oscillate. Specifically, $V_\theta$ oscillates around a zero mean at a frequency about $25\%$ of that of $V_y$. By contrast, $V_r$ exhibits smaller-amplitude oscillations at a frequency close to that of $V_y$. Unlike $V_\theta$, which oscillates around zero throughout the evolution, $V_r$ remains positive over the interval $225\lesssim T\lesssim600$. These behaviours are consistent with the weakly three-dimensional trajectory of this case shown in figure~\ref{fig:3DC_path}.

\begin{figure}
\centerline{\includegraphics[scale=0.65]{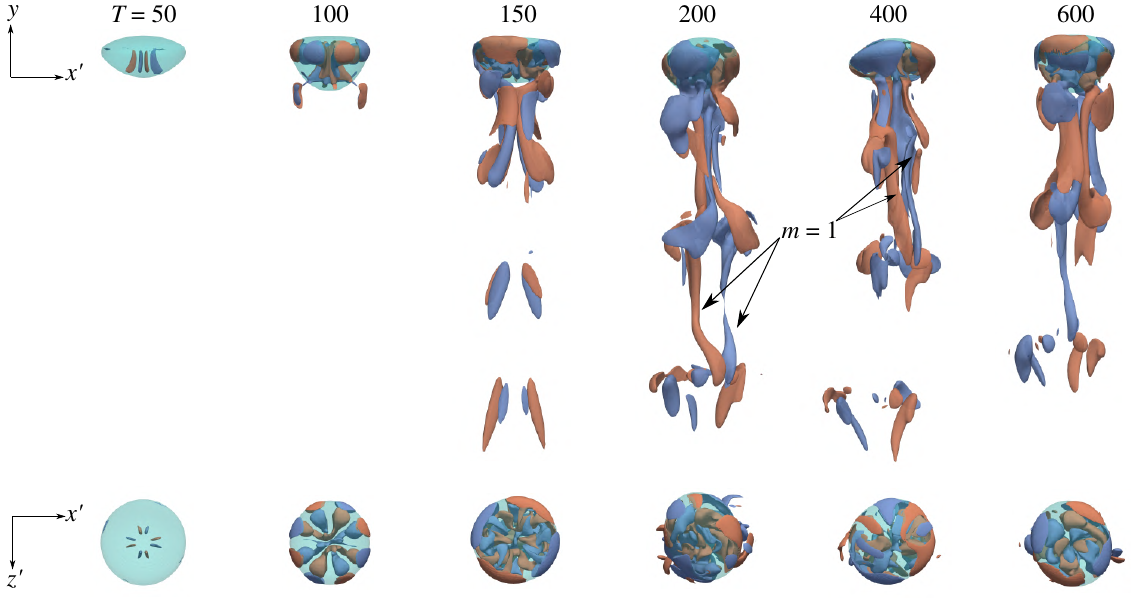}}
\caption{Isosurfaces of the axial vorticity, $\overline{\omega}_y=\pm0.75$, for the droplet with $R=2.4\,\text{mm}$ at selected time instants, indicated by numbers at the top of each panel and marked by circles in figure~\ref{fig:motion_r2.4}. First row: front view; second row: top view. The plane $(x^\prime,z^\prime)$ is horizontal, with the $x^\prime$ and $z^\prime$ axes selected arbitrarily.}
\label{fig:vor_3dc}
\end{figure}

Figure~\ref{fig:motion_r2.4}$(c)$ shows the evolution of the modal energies $\overline{\mathcal{E}}_m$ associated with the various azimuthal modes. These results, together with the vortical structures shown at selected instants in figure~\ref{fig:vor_3dc}, provide insight into the development of the flow topology during the rise of the droplet. According to figure~\ref{fig:motion_r2.4}$(c)$, the first axisymmetry breaking is associated with an azimuthal mode of wavenumber $m=4$. This is remarkable, since in all regimes discussed so far this mode remains stable during the initial stages of axisymmetry breaking. The hallmark of this mode is the presence of four pairs of streamwise vortices, as confirmed by the vortical structures at $T=50$ and $T=100$ in figure~\ref{fig:vor_3dc}. These isosurfaces also suggest that the initial growth of the disturbance is confined within the droplet, as in the previous regimes.

As $T$ increases beyond approximately $70$, the $m=2$ mode, which is the first unstable mode in the $\text{SV}_{RW}$ regime, also becomes unstable. Its modal energy grows in time and reaches a level comparable to that of the $m=4$ mode by $T\approx150$. The corresponding vortical structure indicates that, while these two even modes coexist inside the droplet, the $m=2$ mode dominates in the external vicinity of the droplet and in the far wake. As $T$ increases beyond $150$, two additional azimuthal modes, with $m=1$ and $m=3$, also become unstable. The emergence of the $m=1$ mode is particularly important, since this mode is known to generate a net lift force \citep{fabre2008bifurcations,2024_Shi_drop} and is therefore responsible for the lateral migration of the droplet at later stages. Its signature may be identified in the vortical structures at $T=200$ and $T=400$, where a pair of counter-rotating streamwise vortices is visible in the far wake. However, such an elongated wake does not persist, since the periodic vortex shedding tends to erase the near-wake structure during each cycle of the $V_y$ oscillation, and hence of the shape oscillation. Unlike in a $\text{SV}_{RW}$ scenario, vortex shedding persists even at long time, as illustrated by the snapshot at $T=600$, and no stable spiral organisation of the wake is established.

Figure~\ref{fig:motion_r2.4}$(d)$ shows the evolution of the unwrapped phase angle, $\Phi_m^{\text{U}}$. Except for the $m=2$ mode, all modes remain almost stationary, especially for $T\gtrsim400$, where their phase angles are nearly time-independent. For the $m=2$ mode, the phase angle decreases approximately linearly from $50$ to $-200$ as $T$ increases from $150$ to $400$, yielding an angular velocity $\Omega_2\approx-0.5$, about $67\%$ of that obtained in the terminal state for $R=2.1\,\text{mm}$. However, the coexistence of several modes, together with persistent vortex shedding, prevents the formation of a coherent spiral wake even during this intermediate stage.

Although not shown here, the features described above for $R=2.4\,\text{mm}$ are also observed up to $R=2.6\,\text{mm}$. For the two larger droplets, $R\geq2.8\,\text{mm}$, most of these features remain, in particular the sequence in which the different azimuthal modes become unstable. 
The main difference concerns the dependence of the rise-speed oscillation frequency on $R$. For the smaller droplets in the 3DC regime, typically $R=2.3$ and $2.4\,\text{mm}$, this frequency is close to that obtained in the corresponding axisymmetric configuration. In the axisymmetric configuration, however, it decreases with increasing $R$, from $\overline{f} \approx 0.19$ at $R=2.3\,\text{mm}$ to $\overline{f} \approx 0.15$ at $R=3.0\,\text{mm}$. By contrast, in the fully three-dimensional simulations, $\overline{f}$ remains close to $0.19$ and depends only weakly on $R$. As a result, for $R\geq2.8\,\text{mm}$, the rise-speed oscillation frequency in the three-dimensional case becomes sizeably larger than its axisymmetric counterpart. This suggests that, for sufficiently large droplets, three-dimensional effects play an important role in setting the frequency of the shape oscillations.

\section{Influence of initial disturbances} \label{sec:inf_dis}

We are now in a position to examine how the flow system responds to finite-amplitude initial disturbances, such as those that may be introduced during the experimental release of the droplet. In all simulations discussed in the previous section, the droplet was released from an undisturbed initial condition, with a spherical interface and no imposed finite-amplitude disturbance. Consequently, these simulations alone cannot determine whether some of the transitions identified above are subcritical, in which case more than one stable equilibrium state may be reached depending on the initial conditions. Evidence of such subcritical behaviour has already been reported in previous experimental and numerical studies \citep{wegener2009einfluss,2015_Engberg,charin2019dynamic}: at $R=1.5\,\text{mm}$, the droplet may ultimately rise at two different terminal speeds depending on its initial shape.

\subsection{Initially disturbed states and overview of the results}\label{sec:init_states}

To make this assessment feasible, we consider three asymmetric \q{initial} states, involving both the droplet shape and the disturbance flow field. These states are extracted from transient or fully developed results obtained for the same liquid--liquid system.

The first initial state is taken from the simulation starting from an undisturbed state at $R=1.4\,\text{mm}$, at the instant $T=68$. The corresponding vortical structure is shown in figure~\ref{fig:vor_so}. At this instant, the asymmetry is dominated by the azimuthal mode $m=2$; see figure~\ref{fig:v_rise_so}$(d)$. This state is therefore referred to hereafter as the $m=2$ initial state.

The second initial state corresponds to the terminal state obtained for $R=1.6\,\text{mm}$ when the simulation is initialised from the $m=2$ state defined above. As shown below, using such an $m=2$-disturbed initial state allows the SO path to persist up to $R=1.6\,\text{mm}$, instead of $R=1.45\,\text{mm}$ when the droplet is released from an undisturbed state. In this terminal state, the ratio of modal energies is $\overline{\mathcal{E}}_1/\overline{\mathcal{E}}_2\approx1.2$, indicating that the $m=1$ mode is dominant. We therefore refer to this state as the $m=1$ initial state, although the $m=2$ mode still retains a sizeable amplitude.

The third initial state is the fully developed state obtained at $R=2.2\,\text{mm}$ in a simulation initialised from the undisturbed state. This case belongs to the $\text{SV}_{RW}$ regime. We therefore refer to this state hereafter as the RW initial state, since the corresponding flow structure, shown in figure~\ref{fig:l2_sum}$(d)$, is consistent with a rotating-wave solution. For comparison, the undisturbed initial condition used in the previous section is referred to hereafter as the $m=0$ initial state.

\begin{figure}
\centerline{\includegraphics[scale=0.65]{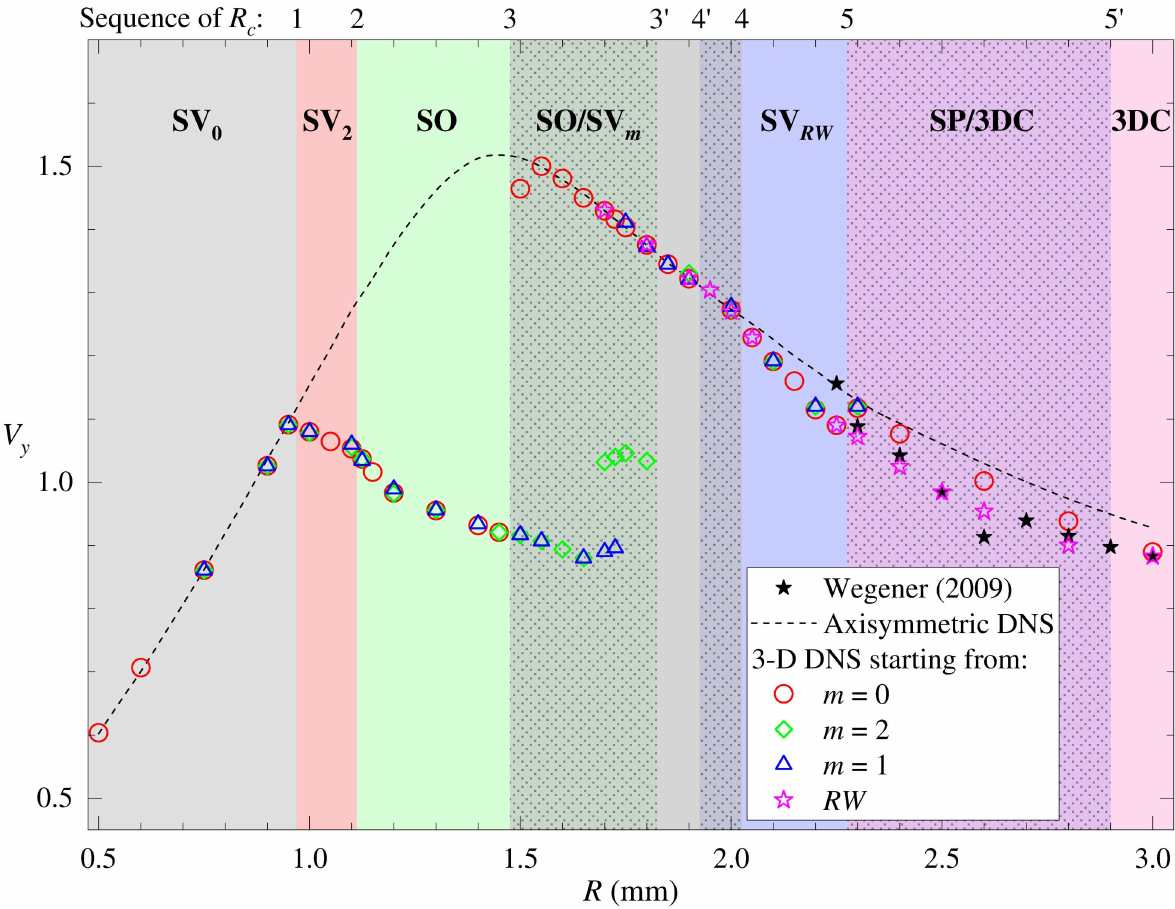}}
\caption{Terminal rise speed of the droplet as a function of the droplet radius $R$ for simulations starting from different initial states. The black dashed line denotes the results obtained in the axisymmetric configuration. Different symbols correspond to fully three-dimensional DNS initialised from different states. The dotted, diamond-patterned regions indicate the multistable intervals of $R$ over which distinct terminal states may be reached at the same $R$ depending on the initial condition.}
\label{fig:v_vs_R_3d_dis}
\end{figure}

To reduce the computational cost, simulations initialised from the $m=2$ and $m=1$ states were carried out for $0.75\,\text{mm}\leq R\leq2.4\,\text{mm}$, whereas those initialised from the $RW$ state were performed for $1.7\,\text{mm}\leq R\leq3.0\,\text{mm}$. Technically, these disturbed initial conditions are generated by restarting the computation from a stored snapshot of the transient three-dimensional flow field. More precisely, the droplet shape and the corresponding velocity and pressure fields obtained at a reference radius $R_s$ are used as the initial condition for a new simulation performed at the target radius $R$. The volume-fraction, velocity and pressure fields are retained in dimensionless form, while the governing dimensionless parameters, in particular $Ga$ and $Bo$, are reset to the values corresponding to the target radius. In physical terms, this procedure amounts to using the shape and the associated internal and external disturbance flow extracted from one droplet size as a finite-amplitude initial perturbation for another droplet size. This differs from earlier numerical studies in which initial-condition effects were mainly explored through the initial droplet shape or orientation \citep{2015_Engberg,charin2019dynamic}; here, the entire three-dimensional disturbance flow is prescribed together with the droplet shape. The procedure is therefore not intended to reproduce a specific experimental release process, but rather to probe the sensitivity of the long-time dynamics to finite-amplitude disturbances with prescribed spatial structure.

Figure~\ref{fig:v_vs_R_3d_dis} shows the terminal rise speeds obtained from these different initial conditions. These results, combined with the characteristic flow structures identified from the azimuthal Fourier analysis, reveal the existence of eight critical droplet radii at which the flow structure may change. Five of them have already been identified from the three-dimensional DNS initialised from the $m=0$, i.e. undisturbed, state. These are $R_{c1}\approx0.97\,\text{mm}$ for the $\text{SV}_0$--$\text{SV}_2$ transition, $R_{c2}\approx1.113\,\text{mm}$ for the $\text{SV}_2$--SO transition, $R_{c3}\approx1.475\,\text{mm}$ for the SO--$\text{SV}_m$ transition, $R_{c4}\approx2.025\,\text{mm}$ for the $\text{SV}_m$--$\text{SV}_{RW}$ transition, and $R_{c5}\approx2.275\,\text{mm}$ for the $\text{SV}_{RW}$--3DC transition. All these critical values are obtained by linear interpolation between neighbouring data points, except for $R_{c1}$, which is estimated using the empirical criterion for the onset of the first internal flow bifurcation proposed by \citet{2025_Shi_drop}; see \S~5.2 therein for details.

The three series of initially disturbed simulations reveal three multistable regimes, highlighted by the dotted, diamond-patterned regions in figure~\ref{fig:v_vs_R_3d_dis}. These regimes are associated with three additional critical droplet radii, denoted by $R'_{c3}$, $R'_{c4}$ and $R'_{c5}$, whose values are $R'_{c3}\approx1.825\,\text{mm}$, $R'_{c4}\approx1.925\,\text{mm}$ and $R'_{c5}\approx2.9\,\text{mm}$, respectively. As $R$ increases, the first multistable regime encountered is $R_{c3}<R<R'_{c3}$, within which the system may evolve towards either the SO branch or the $\text{SV}_m$ branch, depending on the initial conditions. The second multistable regime corresponds to $R'_{c4}<R<R_{c4}$, where the system may evolve towards either the $\text{SV}_0$ or the $\text{SV}_{RW}$ branch. The third one corresponds to $R_{c5}<R<R'_{c5}$, where the ultimate trajectory may be either three-dimensional chaotic or spiralling. The latter state is denoted by SP and is characterised by a clear spiral topology in the near wake, similar to that observed in the $\text{SV}_{RW}$ regime.

These results also clarify the nature of the transitions encountered as $R$ increases. Within the range of finite-amplitude disturbances examined here, the SO--$\text{SV}_m$ transition at $R_{c3}$ and the transition between the axisymmetric vertical branch and the $\text{SV}_{RW}$ branch at $R_{c4}$ behave as subcritical transitions: over finite intervals of $R$, distinct terminal states are obtained depending on the initial condition. Around $R_{c5}$, the transition from the $\text{SV}_{RW}$ branch to the SP branch has the character of a secondary supercritical bifurcation. A weak spiralling motion is already visible for droplet radii smaller than $R_{c5}$; see, for instance, the three-dimensional path at $R=2.2\,\text{mm}$ shown in figure~\ref{fig:3DC_path_dis}. For $R>R_{c5}$, however, the coexistence of the SP and 3DC states shows that the selected long-time dynamics still depends on the initial condition.

By contrast, the $\text{SV}_0$--$\text{SV}_2$ and $\text{SV}_2$--SO transitions show no such initial-condition dependence in the present tests. Near the corresponding thresholds, the same terminal states are reached when the initial condition is switched among the $m=0$, $m=1$ and $m=2$ states, indicating that these two transitions are supercritical within the range of disturbances considered here. The behaviour of the $\text{SV}_0$--$\text{SV}_2$ transition is consistent with the first axisymmetry breaking of a uniform flow past a fixed spherical droplet investigated in our previous study \citep{2025_Shi_drop}. The behaviour of the $\text{SV}_2$--SO transition is, however, somewhat unexpected, since in the corresponding fixed-droplet study the transition from a biplanar-symmetric flow to a uniplanar-symmetric flow was found to be subcritical. This discrepancy may be due to the different constraints imposed in the two configurations. In the fixed-droplet problem, the droplet is spherical, hence non-deformable, and the rise speed is prescribed independently of the drag force. In the freely rising configuration considered here, by contrast, the rise speed decreases in response to the bifurcation, and the droplet undergoes a time-dependent deformation in response to changes in $V_y$. A definitive classification of these transitions would nevertheless require a dedicated stability analysis of the corresponding solution branches.

In the following subsections, we discuss in more detail the dynamical response of the system in the three multistable regimes.

\subsection{The $\text{SO}$/$\text{SV}_m$ regime} \label{sec:regime_bis1}

\begin{figure}
\centerline{\includegraphics[scale=0.65]{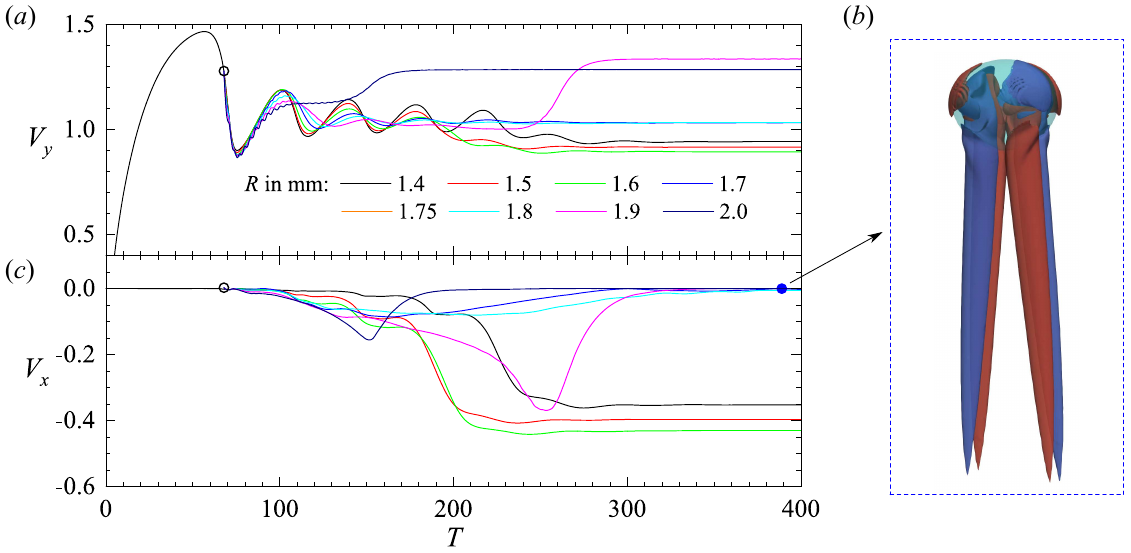}}
\caption{Evolution of $(a)$ the vertical velocity and $(c)$ the horizontal velocity of the droplet for $1.5\,\text{mm}\leq R\leq2.0\,\text{mm}$. All simulations are initialised from the $m=2$ initial state. In both panels, black lines show the evolution in the reference case, namely the case $R=1.4\,\text{mm}$ initialised from the $m=0$ state; the black circle indicates the instant at which this reference result is used to initialise the other simulations. Panel $(b)$ shows isosurfaces of the axial vorticity, $\overline{\omega}_y=\pm0.75$, for the droplet with $R=1.7\,\text{mm}$ at $T=390$, marked by a blue circle in $(c)$.}
\label{fig:v_vor_bis1_m2}
\end{figure}

Figures~\ref{fig:v_vor_bis1_m2}$(a)$ and $(c)$ show the evolution of the vertical and horizontal velocities, respectively, for droplets with $1.5\,\text{mm}\leq R\leq2.0\,\text{mm}$. All simulations are initialised from the $m=2$ state. Remarkably, for $R=1.5$ and $1.6\,\text{mm}$, the droplets acquire a finite horizontal velocity in the terminal state, indicating that they now follow an SO path, as observed for the reference case $R=1.4\,\text{mm}$. For $1.7\,\text{mm}\leq R\leq1.8\,\text{mm}$, the terminal path remains vertical, with $V_x\approx0$, but the terminal rise speed is significantly smaller than in the corresponding axisymmetric configuration. Examination of the terminal vortical structure, shown in figure~\ref{fig:v_vor_bis1_m2}$(b)$ for $R=1.7\,\text{mm}$, reveals that the terminal flow remains biplanar-symmetric. In other words, the system remains on the $m=2$ branch selected by the initial disturbance. Finally, for $R=1.9$ and $2.0\,\text{mm}$, the flow ultimately recovers its axisymmetry, so that $V_x\approx0$ and $V_y$ coincides with its axisymmetric counterpart.

\begin{figure}
\centerline{\includegraphics[scale=0.65]{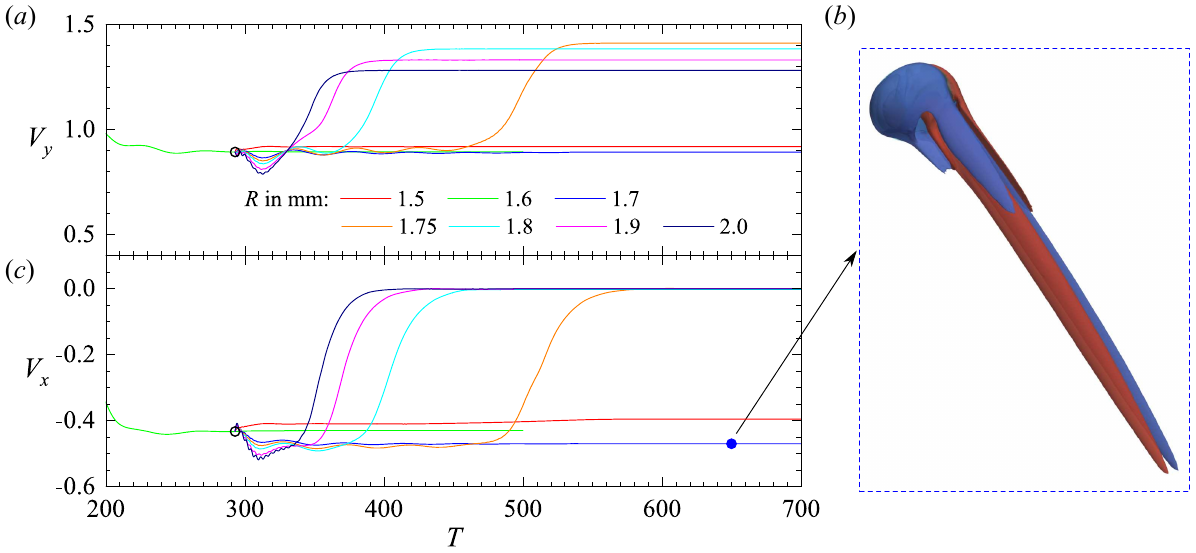}}
\caption{Same as figure~\ref{fig:v_vor_bis1_m2}, but with all simulations initialised from the $m=1$ state. This state corresponds to the transient result obtained for $R=1.6\,\text{mm}$, starting from the $m=2$ initial state, at $T=293$, indicated by an open black circle in both panels. Panel $(b)$ shows isosurfaces of the axial vorticity, $\overline{\omega}_y=\pm0.75$, for the droplet with $R=1.7\,\text{mm}$ at $T=650$, marked by a blue circle in $(c)$.}
\label{fig:v_vor_bis1_m1}
\end{figure}

Figure~\ref{fig:v_vor_bis1_m1} shows the same evolution as figure~\ref{fig:v_vor_bis1_m2}, but with all simulations now initialised from the $m=1$ state. At $R=1.5\,\text{mm}$, the terminal vertical and horizontal velocities do not differ from those obtained from the $m=2$ initial state, indicating that, for this droplet size, the terminal state is not very sensitive to the form of the initial asymmetric disturbance. At $R=1.7\,\text{mm}$, by contrast, the vertical velocity stabilizes at a value approximately $11\%$ smaller than that obtained from the $m=2$ initial state. Moreover, the horizontal velocity grows and stabilizes at a value about half that of $V_y$, whereas it remains vanishingly small throughout the evolution of the same case initialised from the $m=2$ state. The corresponding terminal $\overline{\omega}_y$ isosurfaces, shown in figure~\ref{fig:v_vor_bis1_m1}$(b)$, make clear that the flow is now uniplanar-symmetric, thereby retaining the symmetry of the initial state. Finally, for $R\geq1.75\,\text{mm}$, the flow remains uniplanar-symmetric over a finite time interval whose duration decreases as $R$ increases, before eventually recovering axisymmetry.

The tests above confirm that, for droplets with $1.5\,\text{mm}\leq R\leq1.8\,\text{mm}$, more than one stable equilibrium state may coexist. In particular, for $R=1.7\,\text{mm}$, the terminal flow is axisymmetric when the simulation is initialised from the $m=0$ state, biplanar-symmetric when initialised from the $m=2$ state, and uniplanar-symmetric when initialised from the $m=1$ state. Although not shown in figures~\ref{fig:v_vor_bis1_m2} and \ref{fig:v_vor_bis1_m1}, the same behaviour persists for $R=1.725\,\text{mm}$, confirming that this triple-stable behaviour is not an isolated case.

\subsection{The $\text{SV}_0$/$\text{SV}_{RW}$ regime} \label{sec:regime_bis2}

\begin{figure}
\centerline{\includegraphics[scale=0.65]{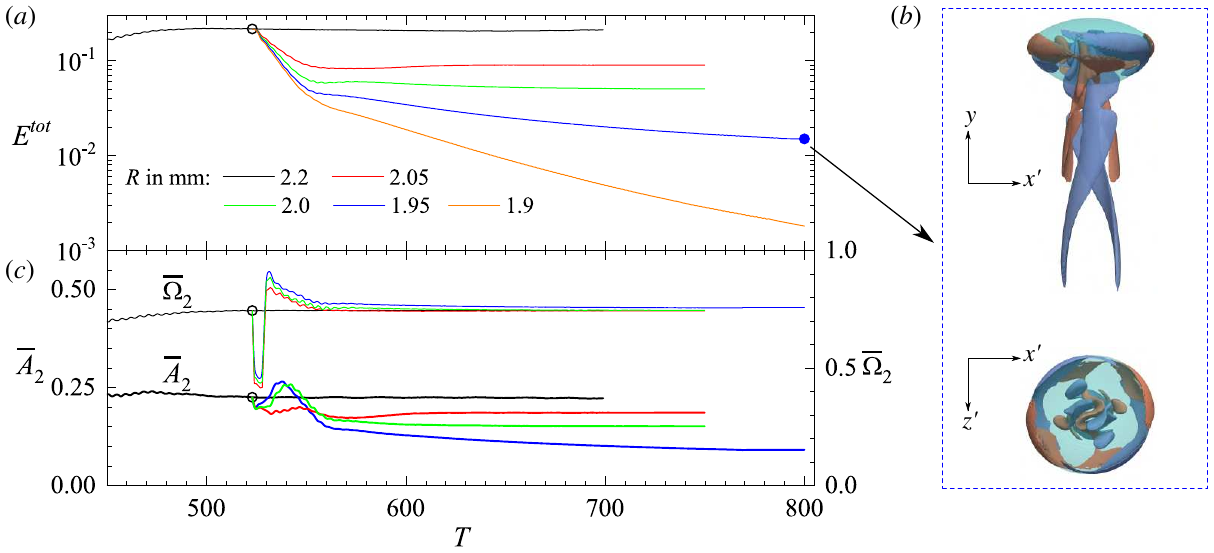}}
\caption{Evolution of quantities characterising the response of droplets initialised from an $RW$ state. $(a)$ Total azimuthal energy $E^{tot}$. $(b)$ Isosurfaces of the axial vorticity, $\overline{\omega}_y=\pm0.5$, for $R=1.95\,\text{mm}$ at $T=800$, marked by a blue circle in $(a)$. $(c)$ Global modal amplitude $\overline{A}_2$ (thick lines, left axis) and angular velocity $\overline{\Omega}_2$ (thin lines, right axis) of the disturbance associated with azimuthal wavenumber $m=2$. In $(a)$ and $(c)$, black lines show the reference case, namely $R=2.2\,\text{mm}$ initialised from the $m=0$ state; the black circle indicates the instant at which this reference result is used to initialise the other simulations.}
\label{fig:v_vor_bis2_rw}
\end{figure}

For $1.8\,\text{mm}<R<2.05\,\text{mm}$, the flow in the simulations initialised from either the $m=2$ or the $m=1$ state always recovers axisymmetry; this is indicated by the decay of the total azimuthal energy, $E^{tot} = E^i + E^e$, to vanishingly small values (not shown).
This is no longer the case when the simulations are initialised from an $RW$ state. Figure~\ref{fig:v_vor_bis2_rw}$(a)$ shows the corresponding evolution of $E^{tot}$ for $1.9\,\text{mm}\leq R\leq2.05\,\text{mm}$. At $R=1.9\,\text{mm}$, $E^{tot}$ rapidly decays to a negligible value. By contrast, for $R\geq1.95\,\text{mm}$, it ultimately saturates at a finite value, increasing from approximately $0.015$ at $R=1.95\,\text{mm}$ to approximately $0.089$ at $R=2.05\,\text{mm}$. The terminal value reached at $R=2.05\,\text{mm}$ agrees with that obtained from the simulation initialised from the $m=0$ state, indicating that the corresponding conditions lie beyond the supercritical side of the transition.

For the two disturbed cases at $R=1.95$ and $2.0\,\text{mm}$, the terminal state corresponds to an $RW$ scenario, similar to that observed for $R=2.05\,\text{mm}$. As a first check, figure~\ref{fig:v_vor_bis2_rw}$(b)$ shows the vortical structure in the fully developed state for $R=1.95\,\text{mm}$. It resembles that observed for $R=2.1\,\text{mm}$ (see figure~\ref{fig:vor_sv_rw}), except that the downstream extension of the wake is shorter. The top view reveals that the disturbance is dominated by the $m=2$ mode. This is further confirmed by examining the evolution of the modal energies up to $m=3$. For both droplets, only $\overline{\mathcal{E}}_2$ retains a finite value in the terminal state, as in the case $R=2.1\,\text{mm}$ shown in figure~\ref{fig:mode_sv_rw}$(b)$.

Figure~\ref{fig:v_vor_bis2_rw}$(c)$ shows the evolution of the global modal amplitude $\overline{A}_2$ and angular velocity $\overline{\Omega}_2$ associated with the $m=2$ mode for $1.95\,\text{mm}\leq R\leq2.05\,\text{mm}$. These quantities are defined in Eqs.~\eqref{eq:global_Qm}--\eqref{eq:global_omega}. For all cases, both eventually approach nearly time-independent values, consistent with the $RW$ state identified for $R=2.1\,\text{mm}$ in \S~\ref{sec:regime_sv_rw}. Combining these results with those of \S~\ref{sec:regime_sv_rw}, we find that $\overline{\Omega}_2\approx0.75$ throughout the $\text{SV}_{RW}$ regime, with only a weak dependence on $R$ over the range $1.95\,\text{mm}\leq R\leq2.25\,\text{mm}$.

\subsection{The $\text{SP}$/3DC regime} \label{sec:regime_bis3}

For $R\geq2.3\,\text{mm}$, the rise path of droplets initialised from an $RW$ state ultimately evolves towards a spiralling motion, hereafter referred to as SP, similar to that observed for bubbles and particles in highly inertial regimes \citep{2012_Ern,2018_Auguste,2025_Shi}. As in the $\text{SV}_0$/$\text{SV}_{RW}$ regime, the selection of this SP branch depends closely on the form of the initial disturbance. Specifically, when the simulation is initialised from either the $m=2$ or the $m=1$ state, the terminal path remains three-dimensional chaotic, as it does for the initially undisturbed case, and the mean terminal rise speed is close to that obtained in the latter case (see figure~\ref{fig:v_vs_R_3d_dis}).

The experimental data of \citet{wegener2009einfluss}, also shown in figure~\ref{fig:v_vs_R_3d_dis}, exhibit a similar trend. In the size range where the SP and 3DC states coexist, most experimental points are closer to the branch obtained from the $RW$ initial state than to that obtained from the undisturbed initial condition. The only clear exception is the point near $R=2.6\,\text{mm}$, for which the measured rise speed is substantially lower. This agreement suggests that the experimental release conditions may favour the selection of an SP-like state, possibly through finite-amplitude disturbances generated during the early transient.

\begin{figure}
\centerline{\includegraphics[scale=0.65]{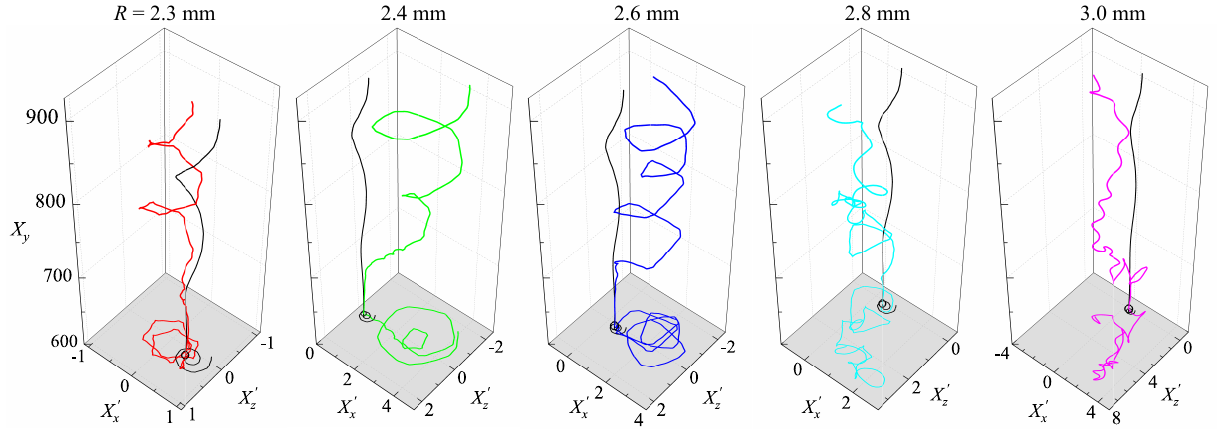}}
\caption{Three-dimensional paths for $2.3\,\text{mm}\leq R\leq3.0\,\text{mm}$ obtained from simulations initialised from an $RW$ state. In all panels, black lines show the evolution of the reference case, namely $R=2.2\,\text{mm}$ initialised from the $m=0$ state; the black circle indicates the instant at which this reference state is used to initialise the other simulations.}
\label{fig:3DC_path_dis}
\end{figure}

Figure~\ref{fig:3DC_path_dis} shows the three-dimensional paths for $2.3\,\text{mm}\leq R\leq3.0\,\text{mm}$. In the reference case, namely $R=2.2\,\text{mm}$ initialised from the $m=0$ state, a weak spiralling structure can already be observed at late stages, for $X_y>600$. The helix diameter increases in time but remains smaller than $0.5R$ during the last turn, which is why this path is still classified as an $\text{SV}_{RW}$ state. For simulations initialised from an $RW$ state, a clear spiralling structure already appears at $R=2.3\,\text{mm}$. As $R$ increases, the typical helix diameter does not vary monotonically: it increases from approximately $1.5R$ at $R=2.3\,\text{mm}$ to $3R$ at $R=2.4\,\text{mm}$, before decreasing to $2R$ at $R=2.6\,\text{mm}$. At larger $R$, this spiralling structure coexists with a 3DC path at $R=2.8\,\text{mm}$ and eventually disappears at $R=3.0\,\text{mm}$, where the path remains of 3DC type. The spiralling paths observed here differ from those reported for freely moving bubbles \citep{mougin2002path,2016_Cano-Lozano,2025_Shi} and solid particles \citep{veldhuis2009freely,2018_Auguste} in that their primary path frequency is much lower. In terms of the corresponding Strouhal number, the latter two systems typically yield $0.075\lesssim St\lesssim0.15$. In the present SP regime, the helix pitch is about $100R$ for $2.3\,\text{mm}\leq R\leq2.6\,\text{mm}$, corresponding to $St\approx0.02$, i.e. about one fifth of the typical values reported in those systems.

\begin{figure}
\centerline{\includegraphics[scale=0.65]{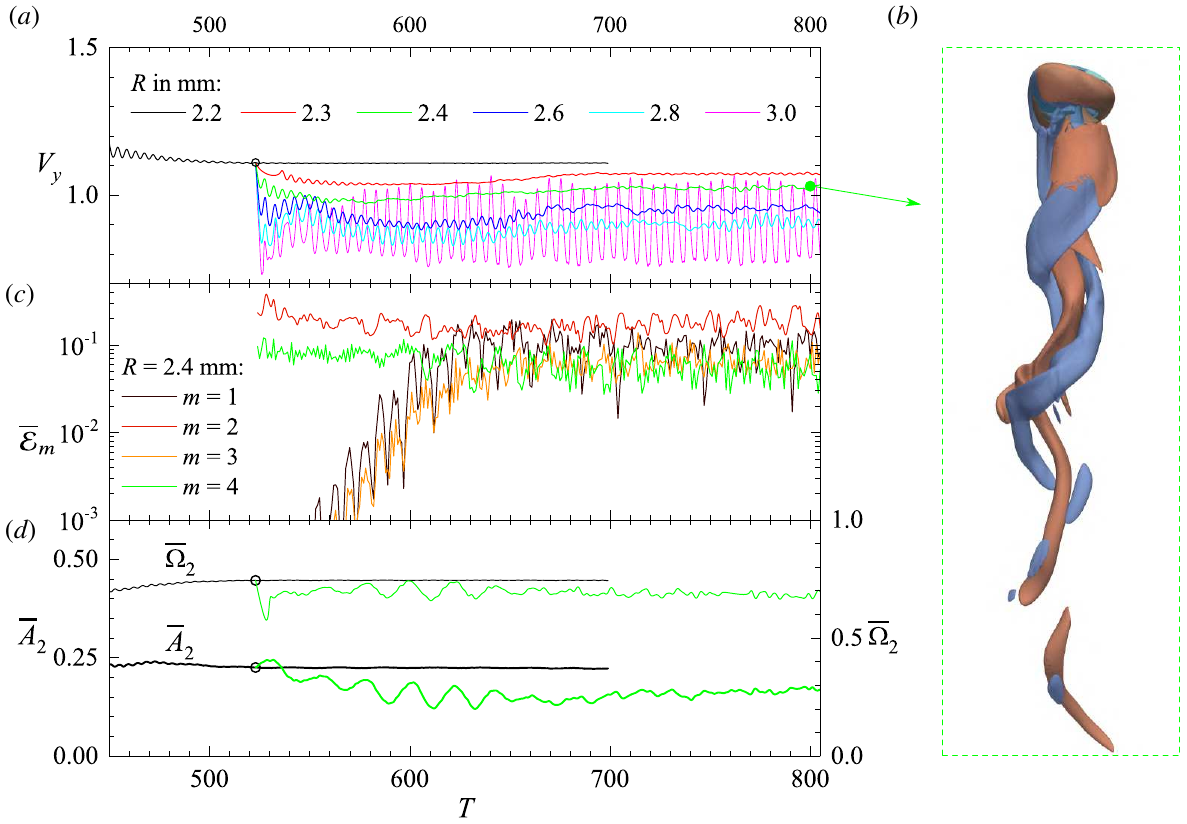}}
\caption{$(a)$ Evolution of the droplet rise speed for $2.3\,\text{mm}\leq R\leq3.0\,\text{mm}$. $(b)$ Isosurfaces of the axial vorticity, $\overline{\omega}_y=\pm0.75$, for $R=2.4\,\text{mm}$ at $T=800$, marked by a green circle in $(a)$. $(c)$ Evolution of the normalised modal energies for $R=2.4\,\text{mm}$. $(d)$ Evolution of the global modal amplitude $\overline{A}_2$ (thick line, left axis) and angular velocity $\overline{\Omega}_2$ (thin line, right axis) of the disturbance associated with azimuthal wavenumber $m=2$ for $R=2.4\,\text{mm}$. All simulations are initialised from the $RW$ state.}
\label{fig:v_vor_bis3_rw}
\end{figure}

Figure~\ref{fig:v_vor_bis3_rw}$(a)$ shows the evolution of the rise speed. For droplets up to $R=2.8\,\text{mm}$, for which an SP path is observed, $V_y$ ultimately reaches a weakly oscillatory state. The relative amplitude of these oscillations, measured with respect to the late-time mean value, increases from less than $1\%$ for $R=2.3\,\text{mm}$ to about $3.5\%$ for $R=2.8\,\text{mm}$. Remarkably, both the mean rise speed and the oscillation amplitude are smaller than those obtained when the simulation is initialised from the $m=0$ state. For instance, at $R=2.4\,\text{mm}$, the mean value and relative oscillation amplitude of $V_y$ are approximately $1.02$ and $0.8\%$, respectively, whereas the corresponding values are $1.08$ and $12\%$ when the simulation is initialised from the $m=0$ state (see figure~\ref{fig:motion_r2.4}$a$).

To understand this difference, we examine the flow structure and modal content for $R=2.4\,\text{mm}$. Figure~\ref{fig:v_vor_bis3_rw}$(b)$ shows the corresponding vortical structure in the fully developed state. The wake displays a clear spiral topology, similar to that observed in the $\text{SV}_{RW}$ regime, although the spiralling motion is now accompanied by a finite lateral displacement of the droplet. The modal-energy evolution shown in figure~\ref{fig:v_vor_bis3_rw}$(c)$ indicates that the $m=2$ mode dominates the saturated state, whereas the other modes remain much weaker. This contrasts with the 3DC state obtained from the $m=0$ initial condition at the same droplet size, for which several azimuthal modes coexist and persistent vortex shedding prevents the establishment of a coherent spiral wake (see figure~\ref{fig:motion_r2.4}$(c)$ and figure~\ref{fig:vor_3dc}).

Figure~\ref{fig:v_vor_bis3_rw}$(d)$ shows the evolution of the global modal amplitude $\overline{A}_2$ and angular velocity $\overline{\Omega}_2$ associated with the $m=2$ mode for $R=2.4\,\text{mm}$. After a transient stage, both quantities approach nearly time-independent values, indicating that this SP state retains the rotating-wave character of the initial disturbance. The same behaviour was found for the other droplets following an SP path, although these results are not shown here. Unlike in the $\text{SV}_{RW}$ regime, where $\overline{\Omega}_2$ depends only weakly on $R$, the mean value of $\overline{\Omega}_2$ in the SP regime decreases from approximately $0.71$ at $R=2.3\,\text{mm}$ to approximately $0.54$ at $R=2.8\,\text{mm}$. This indicates that the axial rotation of the disturbance gradually \q{slows down} as $R$ increases.

\section{Summary and outlook} \label{sec:conclusion}

\subsection{Main findings}

We carried out a series of three-dimensional DNS to clarify the physical mechanisms governing the motion of a single deformable droplet rising freely in a large body of an immiscible liquid. Since the present study is primarily concerned with liquid--liquid systems in which the droplet viscosity is comparable to that of the surrounding fluid, we focused on toluene droplets rising in clean, i.e. surfactant-free, water, for which the drop-to-fluid viscosity ratio is $\mu^\ast=0.62$. The droplet radius was varied in the range $0.5\,\text{mm}\leq R\leq3.0\,\text{mm}$. For this toluene--water system, the simulations show that the first departure from flow axisymmetry occurs while the droplet still rises vertically, through an $m=2$ disturbance whose growth is first detected inside the droplet. As $R$ is increased further, the droplet successively exhibits steady oblique motion, a return to vertical rise, a rotating-wave vertical state and, finally, fully three-dimensional chaotic paths. In most of the non-axisymmetric regimes, the first visible sign of flow asymmetry is found inside the droplet before a pronounced asymmetry develops in the wake, highlighting the central role of the internal flow instability in shaping the subsequent wake structure and droplet dynamics. These points are discussed in more detail in the following.

As a reference, we first considered an imposed axisymmetric configuration. In quantitative agreement with the axisymmetric predictions of \citet{2011_Baumler} and \citet{engberg2014numerical} for the same liquid--liquid system, the terminal rise speed was found to reach a maximum at $R\approx1.5\,\text{mm}$. By contrast, the aspect ratio, defined as the ratio of the major to minor axes of the droplet, reaches its maximum at a larger droplet size, $R\approx2.1\,\text{mm}$. For $R\gtrsim2.05\,\text{mm}$, the droplet undergoes regular shape oscillations. Expressed in terms of the Strouhal number, the corresponding frequency is $St\approx0.33$, in close agreement with the prediction of \citet{engberg2014numerical}.

We then examined the fully three-dimensional evolution. The main findings may be summarised through the sequence of rise regimes encountered as $R$ increases. When available, we also recall relevant previous observations concerning the droplet rise speed and path.

For $R$ increasing up to approximately $0.97\,\text{mm}$, the droplet follows a steady vertical path and the disturbance flow remains axisymmetric. We denote this regime as $\text{SV}_0$, with the subscript \q{0} emphasizing the axisymmetric nature of the disturbance flow. Since the droplet remains nearly spherical in this regime, its terminal rise speed is also well predicted by drag correlations for spherical droplets, such as that proposed by \citet{2024_Shi_drop}; see figure~\ref{fig:cd_vs_re_2d} for a relevant comparison.

For $0.97\,\text{mm}\lesssim R\leq1.1\,\text{mm}$, the droplet still follows a steady vertical path. However, the terminal disturbance flow is now biplanar-symmetric, characterised by two pairs of counter-rotating streamwise vortices in the wake. The resulting three-dimensional wake increases the drag on the droplet, making the terminal rise speed significantly smaller than in the corresponding axisymmetric configuration, in line with the observations of \citet{wegener2010terminal}, \citet{engberg2014numerical} and \citet{charin2019dynamic}. The azimuthal Fourier analysis of the transient disturbance flow, shown in figure~\ref{fig:mode_sv2}, indicates that the axisymmetry breaking is associated with an azimuthal mode of wavenumber $m=2$. We therefore denote this regime as $\text{SV}_2$. The $\text{SV}_0$--$\text{SV}_2$ transition occurs at a critical radius between $0.95$ and $1.0\,\text{mm}$, in good quantitative agreement with the criterion proposed in our previous study \citep{2025_Shi_drop} for the internal flow instability of a uniform flow past a fixed spherical droplet. Applying that criterion to the liquid--liquid system considered here yields a threshold radius of approximately $0.97\,\text{mm}$. This agreement suggests that the $\text{SV}_0$--$\text{SV}_2$ transition of a freely rising droplet is closely linked to the internal flow instability observed when the same droplet is held fixed in a uniform flow.

For $1.125\,\text{mm}\leq R\leq1.45\,\text{mm}$, the droplet follows a steady oblique path, denoted as SO. The wake still consists of two pairs of counter-rotating streamwise vortices (see figure~\ref{fig:vor_so}). However, one pair is now significantly weaker than the other, so that the flow is only uniplanar-symmetric. This stronger wake asymmetry further enhances viscous dissipation and makes the reduction of the rise speed, compared with the corresponding axisymmetric configuration, more pronounced than in the $\text{SV}_2$ regime. The azimuthal Fourier analysis (see figure~\ref{fig:v_rise_so}$d$) indicates that the initial axisymmetry breaking is still driven by the $m=2$ mode, as in the $\text{SV}_2$ regime. At a later stage, however, the $m=1$ mode also becomes unstable, leading to the $\text{SV}_2$--SO transition. The two modes coexist in the terminal state. This distinguishes the present SO regime from the steady oblique path of a freely moving solid particle, for which only the $m=1$ mode is involved and the first non-vertical path corresponds to a $\text{SV}_0$--SO transition \citep{fabre2012steady}.

For $1.5\,\text{mm}\leq R\leq2.0\,\text{mm}$, the droplet again follows a steady vertical path. We denote this regime as $\text{SV}_m$, where the subscript indicates that different azimuthal modes may appear transiently or, in one case, persist in the terminal state. For all droplets in this range except $R=1.5\,\text{mm}$, the flow ultimately recovers axisymmetry and the terminal rise speed coincides with that obtained in the axisymmetric configuration. The case $R=1.5\,\text{mm}$ is peculiar: the $m=2$ mode first grows weakly and then decays, after which the axisymmetry of the flow breaks down through an azimuthal mode with wavenumber $m=3$. The corresponding terminal state remains vertical, but the rise speed is approximately $4\%$ smaller than its axisymmetric counterpart. The recovery of an almost axisymmetric vertical-rise state for $R>1.5\,\text{mm}$ explains the abrupt increase in the terminal rise speed observed in both experiments and previous simulations around this droplet size.

For $2.05\,\text{mm}\lesssim R\lesssim2.25\,\text{mm}$, the flow axisymmetry breaks down again while the droplet path remains nearly vertical. We denote this regime as $\text{SV}_{RW}$. During the transient stage, the rise speed and droplet shape undergo oscillations similar to those observed in the imposed axisymmetric configuration. These oscillations gradually decay, after which the droplet adopts a frozen, weakly asymmetric shape and rises at a nearly time-independent speed, slightly smaller than its axisymmetric counterpart. The terminal wake exhibits a spiral organisation and is dominated by an azimuthal mode with wavenumber $m=2$. The global modal amplitude $\overline{A}_2$ and angular velocity $\overline{\Omega}_2$ both approach nearly time-independent values, indicating that this state has the character of a rotating-wave solution. The corresponding rotation period is $T_R\approx8.4$, with only a weak dependence on $R$ over the range considered.

For $R\gtrsim2.3\,\text{mm}$, the $\text{SV}_{RW}$ regime is replaced by a fully three-dimensional chaotic regime, denoted as 3DC. In this regime, the rise speed keeps oscillating at long time, with a mean value and a frequency close to those obtained in the corresponding axisymmetric configuration. This indicates that shape oscillations remain a central ingredient of the dynamics. The flow topology is, however, fundamentally three-dimensional. The initial axisymmetry breaking is associated with the $m=4$ mode, followed by the growth of the $m=2$, $m=1$ and $m=3$ modes. The emergence of the $m=1$ mode is particularly important, since it generates a net lateral force and is therefore responsible for the lateral migration of the droplet. In contrast to the $\text{SV}_{RW}$ regime, vortex shedding persists even at long time, preventing the formation of a coherent spiral wake.

The influence of finite-amplitude initial disturbances was then examined by initialising the simulations from three asymmetric states, referred to as the $m=2$, $m=1$ and $RW$ initial states. These tests reveal the existence of three multistable regimes. The first one occurs for $R_{c3}<R<R'_{c3}$, where the system may evolve towards either the SO branch or the $\text{SV}_m$ branch depending on the initial condition. In this range, axisymmetric, biplanar-symmetric and uniplanar-symmetric terminal states may coexist. The second multistable regime occurs for $R'_{c4}<R<R_{c4}$, where the droplet may ultimately follow either the $\text{SV}_0$ branch or the $\text{SV}_{RW}$ branch. The third one occurs for $R_{c5}<R<R'_{c5}$, where the terminal state may be either 3DC or spiralling. The latter, denoted as SP, retains the rotating-wave character of the $RW$ initial disturbance but develops a finite lateral motion. Compared with the 3DC state obtained from the undisturbed initial condition, vortex shedding is strongly suppressed in the SP state, allowing a coherent spiralling wake to persist. These initial-condition tests also provide useful indications regarding the possible nature of the corresponding transitions. The SO--$\text{SV}_m$ transition and the transition between the axisymmetric vertical branch and the $\text{SV}_{RW}$ branch appear consistent with subcritical behaviour, since finite-amplitude disturbances allow the system to reach terminal states that are not obtained from the undisturbed initial condition over the same range of $R$. By contrast, the $\text{SV}_0$--$\text{SV}_2$ and $\text{SV}_2$--SO transitions appear consistent with a supercritical scenario within the range of disturbances examined here. A definitive classification would, however, require a dedicated stability or continuation analysis.

\subsection{Future work}

The present work leaves several open questions concerning the mechanisms underlying the flow and path instabilities of freely rising droplets. 

One of them concerns the role of time-dependent droplet deformation in the intermediate size range $1.0\,\text{mm}\lesssim R\lesssim2.0\,\text{mm}$. For $R\gtrsim2.0\,\text{mm}$, simulations initialised from an undisturbed state reveal a transition from the axisymmetric vertical branch to the $\text{SV}_{RW}$ regime at $R\approx2.05\,\text{mm}$, very close to the onset of shape oscillations observed in the axisymmetric configuration. This suggests that time-dependent deformation may promote the transition towards the $\text{SV}_{RW}$ regime. By contrast, for $R\lesssim1.0\,\text{mm}$, the simulations reveal an $\text{SV}_0$--$\text{SV}_2$ transition at a critical radius between $0.95$ and $1.0\,\text{mm}$, close to the threshold $R\approx0.97\,\text{mm}$ predicted for the first internal flow instability of a uniform flow past a fixed spherical, and hence non-deformable, droplet in the same liquid--liquid system \citep{2025_Shi_drop}. This indicates that time-dependent deformation has little influence on the $\text{SV}_0$--$\text{SV}_2$ transition at small $R$.

It is not possible to draw a firm conclusion in the intermediate size range. Starting from an undisturbed state, the present simulations show that the $\text{SV}_2$--SO transition occurs for $R$ between $1.1$ and $1.125\,\text{mm}$. This threshold appears to be independent of the initial condition, suggesting that the underlying bifurcation is supercritical. In the fixed-droplet configuration considered by \citet{2025_Shi_drop}, a secondary bifurcation from a biplanar-symmetric flow, corresponding to an $\text{SV}_2$-type state, to a uniplanar-symmetric flow, corresponding to an SO-type state, was also observed, but was found to be subcritical. This difference cannot be attributed solely to time-dependent deformation. Indeed, at the $\text{SV}_2$--SO threshold in the present freely rising configuration, the droplet is already non-spherical, with $\upchi\approx1.25$ for $R=1.125\,\text{mm}$ before axisymmetry breaking, whereas the droplet was imposed to be spherical in the fixed-droplet configuration.

As $R$ exceeds about $1.5\,\text{mm}$, the terminal path switches from SO back to steady vertical, accompanied by an abrupt increase in the terminal rise speed. No analogous abrupt transition has been reported in the fixed-droplet configuration, suggesting that droplet deformation may play a role. To test this possibility, we performed additional simulations of a slightly modified liquid--liquid system in which all physical properties were kept unchanged except for the surface tension, which was reduced by about $15\%$. This modification enhances the deformation of the droplet. The results, not shown here, indicate that the abrupt recovery of the vertical path then occurs at an aspect ratio of approximately $1.45$, corresponding to $R=1.7\,\text{mm}$ in the modified system, rather than at the threshold aspect ratio $\upchi\approx1.35$ found in the original system. This suggests that this transition is not governed by a fixed critical aspect ratio. It also indicates that time-dependent deformation may favour the SO state, thereby delaying the transition back to a steady vertical path.

A natural way to clarify the mechanisms involved in these transitions would be to perform a global linear-stability analysis following the methodology developed by \citet{2024_Bonnefis}. However, for the two intermediate transitions discussed here, namely the $\text{SV}_2$--SO and SO--$\text{SV}_m$ transitions, the state immediately preceding the transition is already non-axisymmetric. A global linear-stability analysis of these transitions would therefore have to be performed in a fully three-dimensional setting, making it computationally demanding. A more feasible first step would be to consider the corresponding fixed-droplet problem, as in \citet{2025_Shi_drop}, but with both the three-dimensional droplet shape and the incident flow prescribed from the freely rising state immediately preceding the transition. A similar strategy was adopted by \citet{cano2016global} to examine the influence of shape asymmetry on the first path instability of rising bubbles. Work along this line is currently in progress.

A second open question concerns more general liquid--liquid systems, and in particular the role of the drop-to-fluid viscosity ratio $\mu^\ast$ in selecting the nature of the first path instability \citep{2026_gode}. At one limit, freely rising bubbles may be viewed as weakly viscous droplets, corresponding formally to $\mu^\ast\rightarrow0$. At the threshold of their first path instability, they follow a zigzag path (ZZ) through a Hopf bifurcation driven by an azimuthal mode with wavenumber $m=1$ \citep{mougin2002path,tchoufag2015weakly,2024_Bonnefis}. At the other limit, highly viscous droplets, including solid particles as a limiting case, undergo a steady oblique path through a pitchfork bifurcation, again involving an $m=1$ mode \citep{2004_Jenny,2015_Albert,2018_Auguste}. In both limits, the disturbance flow remains axisymmetric before the first path instability sets in.

The situation observed in the moderate viscosity ratio $\mu^\ast=0.62$ differs from both limiting cases. The first non-vertical path is still steady oblique, as in the highly viscous limit, but it involves an internal $\text{SV}_2$--SO transition rather than a direct, externally driven $\text{SV}_0$--SO transition. In other words, the axisymmetry of the disturbance flow has already broken down before the path instability sets in. How the viscosity ratio controls both the nature of the first non-vertical path and the sequence of flow-structure transitions therefore remains an open problem. Addressing it would require a systematic exploration of liquid--liquid systems over a broad range of $\mu^\ast$, ideally combining fully three-dimensional DNS with global stability analyses of both fixed and freely moving deformable droplets.

\appendix

\section{Azimuthal Fourier analysis in the cross-stream plane}
\label{app:fft}

This appendix provides the numerical details of the azimuthal Fourier analysis used in \S~\ref{sec:regime_sv2} and \S~\ref{sec:regime_sv_rw}. At each time instant, the cross-stream plane is defined as the plane passing through the droplet centroid $\boldsymbol{x}$ and normal to the instantaneous velocity direction $\boldsymbol{e}_v=\boldsymbol{v}/|\boldsymbol{v}|$. Two orthonormal basis vectors, $\boldsymbol{e}_1$ and $\boldsymbol{e}_2$, are introduced in this plane. In the present implementation, $\boldsymbol{e}_1$ is obtained by projecting the laboratory-frame $x^\prime$ direction onto the plane normal to $\boldsymbol{e}_v$ and then normalising the resulting vector. The second in-plane unit vector is defined as
\begin{equation}
\boldsymbol{e}_2=\boldsymbol{e}_v\times\boldsymbol{e}_1 .
\label{eq:app_e2}
\end{equation}
Thus, $(\boldsymbol{e}_1,\boldsymbol{e}_2,\boldsymbol{e}_v)$ forms a local orthonormal frame attached to the instantaneous droplet motion. Any sampling point in the cross-stream plane is then parametrised by polar coordinates $(r,\theta)$ as
\begin{equation}
\boldsymbol{x}_s(r,\theta)=\boldsymbol{x}+r\cos\theta\,\boldsymbol{e}_1+r\sin\theta\,\boldsymbol{e}_2 ,
\label{eq:app_sampling_point}
\end{equation}
where $r$ is the distance from the droplet centroid and $\theta$ is the azimuthal angle measured in the $(\boldsymbol{e}_1,\boldsymbol{e}_2)$ plane.

As stated in \S~\ref{sec:regime_sv2}, the scalar field analysed here is the streamwise vorticity component within the droplet interior. The continuous definitions of the azimuthal Fourier coefficient and the modal energy are given in Eqs.~\eqref{eq:app_qhat_cont} and \eqref{eq:app_Em_cont}. In the numerical implementation, for each selected radius $r_j$, the Fourier coefficient of wavenumber $m$ is evaluated as
\begin{equation}
\hat{q}_m(r_j,t)\approx\frac{1}{N_\theta}\sum_{k=0}^{N_\theta-1}q(r_j,\theta_k,t)\exp(-\mathrm{i}m\theta_k),
\label{eq:app_qhat_disc}
\end{equation}
with
\begin{equation}
\theta_k=\frac{2\pi k}{N_\theta},\qquad k=0,1,\ldots,N_\theta-1 .
\label{eq:app_theta}
\end{equation}
The corresponding modal energy is approximated as
\begin{equation}
\mathcal{E}_m(t)\approx\sum_{j=1}^{N_r}|\hat{q}_m(r_j,t)|^2 r_j\Delta r .
\label{eq:app_Em_disc}
\end{equation}
This quantity is used to identify the dominant azimuthal wavenumber and to monitor its temporal evolution.

In addition to the modal energy, the global complex Fourier coefficient introduced in \S~\ref{sec:regime_sv_rw} is evaluated numerically by radially integrating the complex Fourier coefficients themselves. Its discrete form reads
\begin{equation}
Q_m(t)\approx\sum_{j=1}^{N_r}\hat{q}_m(r_j,t)\,r_j\Delta r .
\label{eq:app_Qm_disc}
\end{equation}
The corresponding global modal amplitude and phase angle are then obtained from $A_m(t)=|Q_m(t)|$ and $\Phi_m(t)=\arg(Q_m(t))$, respectively, as in Eqs.~\eqref{eq:global_Am} and \eqref{eq:global_phase}. When the angular velocity $\Omega_m$ is required, the phase is first unwrapped in time before numerical differentiation according to Eq.~\eqref{eq:global_omega}.

In the present implementation, the radial and azimuthal resolutions are $N_r=40$ and $N_\theta=128$, respectively. The sampled radii are distributed uniformly over the interval $[r_{\min},r_{\max}]=[0.05R,0.95R]$, consistent with the definition used in \S~\ref{sec:regime_sv2}. The values of $N_r$, $N_\theta$, $r_{\min}$ and $r_{\max}$ were found to provide sufficient resolution of the internal flow structure.

\backsection[Acknowledgements]{The computations were carried out on the HPC cluster \emph{hemera} at HZDR. P.S. acknowledges fruitful discussions with B. Scheid, whose valuable comments were helpful in identifying the rotating-wave regime.}

\backsection[Funding]{This work is funded by the Deutsche Forschungsgemeinschaft (DFG, German Research Foundation) (P.S., grant number 501298479). P.S. also acknowledges support from the Fonds de la Recherche Scientifique -- FNRS under Grant n$^\circ$~40031630, during which the present manuscript was prepared.}

\backsection[Declaration of interests]{The authors report no conflict of interest.}

%\backsection[Data availability statement]{The data that support the findings of this study are openly available in [repository name] at http://doi.org/[doi], reference number [reference number]. See JFM's \href{https://www.cambridge.org/core/journals/journal-of-fluid-mechanics/information/journal-policies/research-transparency}{research transparency policy} for more information}

\backsection[Author ORCIDs]{

Pengyu Shi, https://orcid.org/0000-0001-6402-4720; 

Dirk Lucas, https://orcid.org/0000-0003-0463-2278;

Jie Zhang, https://orcid.org/0000-0002-2412-3617;

\'{E}ric Climent, https://orcid.org/0000-0001-9538-338X;

Dominique Legendre, https://orcid.org/0000-0002-6021-7119.
}

%\backsection[Author contributions]{Authors may include details of the contributions made by each author to the manuscript'}

\bibliographystyle{jfm}
\bibliography{jfm}
\end{document}